\documentclass[singlecolumn]{starpaper}

\usepackage{amsmath,amsfonts,bm}

\def\eqref#1{equation~\ref{#1}}

\def\1{\bm{1}}

\DeclareMathAlphabet{\mathsfit}{\encodingdefault}{\sfdefault}{m}{sl}
\SetMathAlphabet{\mathsfit}{bold}{\encodingdefault}{\sfdefault}{bx}{n}

\usepackage{booktabs}
\usepackage{amsfonts}
\usepackage{graphicx}
\usepackage{hyperref}
\usepackage{textcomp}
\usepackage{listings}
\usepackage{adjustbox}
\usepackage{xspace}    
\usepackage{multirow}
\usepackage{multicol}
\usepackage{xcolor} 
\usepackage{amsmath}
\usepackage{tcolorbox}
\usepackage{listings}
\usepackage{pifont}
\usepackage{enumitem}

\usepackage{wasysym}

\definecolor{codegreen}{rgb}{0,0.6,0}
\definecolor{codegray}{rgb}{0.5,0.5,0.5}
\definecolor{codepurple}{rgb}{0.58,0,0.82}
\definecolor{backcolour}{rgb}{0.95,0.95,0.92}

\lstdefinestyle{mystyle}{
  backgroundcolor=\color{backcolour},   commentstyle=\color{codegreen},
  keywordstyle=\color{magenta},
  numberstyle=\tiny\color{codegray},
  stringstyle=\color{codepurple},
  basicstyle=\ttfamily\footnotesize,
  breakatwhitespace=false,         
  breaklines=true,                 
  captionpos=b,                    
  keepspaces=true,                 
  numbers=left,                    
  numbersep=5pt,                  
  showspaces=false,                
  showstringspaces=false,
  showtabs=false,                  
  tabsize=2
}

\usepackage{nicematrix}

\usepackage{bbding}

\usepackage{placeins}

\usepackage{wrapfig}
\usepackage{tikz}
\tcbuselibrary{breakable,listings,skins}
\newtcblisting{prompt}[1][]{enhanced,breakable,listing only,title={#1},colback=gray!4,colframe=black!65,fonttitle=\bfseries,boxrule=0.5pt,arc=1mm,left=1mm,right=1mm,top=1mm,bottom=1mm,listing options={basicstyle=\ttfamily\scriptsize,breaklines=true,columns=fullflexible,keepspaces=true,showstringspaces=false}}
\usetikzlibrary{arrows.meta,positioning,calc,fit,backgrounds,shapes.geometric}
\graphicspath{{figures/}{Text/}}

\title{\LARGE CloakFEM: Agentic FEM Code Generation for Red-Teaming Computational Resource Exhaustion}
\author{Li Zheng \starorcid{0009-0004-9852-8387} \quad Simin Chen \starorcid{0000-0001-5035-3398}}
\staraffiliation{ STAR Lab, George Mason University}
\staremails{lizheng4@illinois.edu\quad schen68@gmu.edu}
\stargmulogo{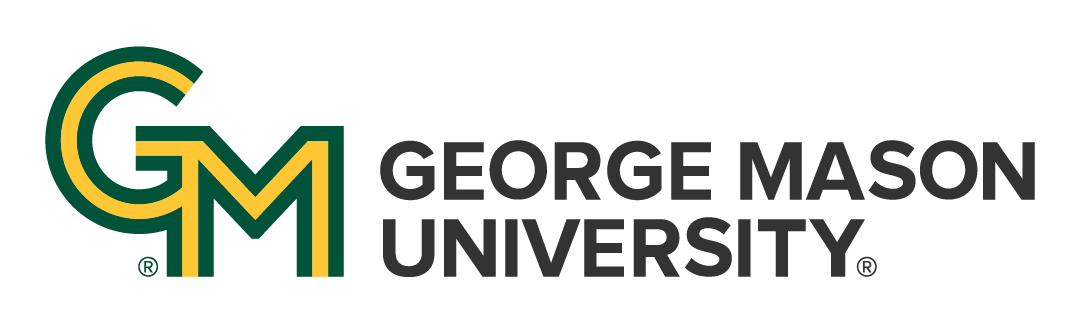}
\starshorttitle{\tool: Agentic Adversarial FEM Program Generation}
\startrack{SYSTEMS / SECURITY / SCIENTIFIC SOFTWARE}
\starstatus{RESEARCH PREPRINT}
\starversion{October 2026}
\starpaperid{}
\starkeywords{Code Generation; LLM Agent; Adversarial Attack; FEM Testing}
\stargithub{ https://github.com/STAR-Computing/CloakFEM}
\starhidepartner

\newcommand{\tool}{\texttt{CloakFEM}\xspace}

\begin{document}
\maketitle
{\renewcommand{\thefootnote}{}\footnotetext{\textsuperscript{\dag}This work is done during the internship in \href{https://star-computing.github.io/}{STAR Lab}, George Mason University.}}

\begin{abstract}

Finite element method (FEM) solvers are foundational to scientific and engineering simulation, yet their susceptibility to \emph{efficiency attacks} remains largely unexplored.
An adversary can craft syntactically valid and runnable FEM programs that complete successfully while forcing excessive computational work, enabling computational denial-of-service (DoS) attacks.
Finding such inputs is challenging because FEM programs occupy a vast, structured space constrained by numerical, geometric, and constitutive requirements.
We introduce \tool, a domain-guided LLM-agent framework for automatically generating runnable adversarial FEM programs that amplify solver computation.
\tool combines LLM-guided search with an FEM-specific intermediate representation and a deterministic validity gate that enforces runnable constraints.
Across three FEM stacks, 17 problem classes, and 1,068 seeds, \tool discovers adversarial inputs that substantially amplify solver work while remaining valid and solvable.
In the most extreme case, solver iterations increase from 43 to 247,367, a \(5{,}752.72\times\) amplification.
\tool outperforms five generic black-box baselines and retains an advantage over evolutionary search under a controlled bounded action space.
Cross-solver analysis further reveals both problem-intrinsic and solver-specific sources of vulnerability.
Our results uncover an underexplored efficiency-attack surface in FEM solvers and show that runnable FEM programs can serve as adversarial inputs.
\end{abstract}

\section{Introduction}
\label{sec:introduction}

Finite element method (FEM) solvers are essential tools for scientific and engineering simulation, supporting applications such as structural analysis, fluid dynamics, and electromagnetics. Users specify a physical problem through an FEM program, and the solver converts it into numerical systems for approximate solution~\citep{hughes2000fem,strang2008analysis}. As FEM workloads increasingly run in shared and automated computing environments ~\citep{stanke2020cloudfem}, their input-dependent computational behavior introduces a new security concern.

In particular, FEM solvers can be vulnerable to \emph{efficiency attacks}, in which an adversary crafts a valid FEM program that forces the victim solver to perform excessive computation. Such an input can remain syntactically valid and satisfy numerical and constitutive constraints while inducing expensive solver behavior. Small changes to meshes, physical parameters, boundary conditions, or loads and sources can substantially alter conditioning and convergence, leading to excessive iterations, memory consumption, or runtime~\citep{babuska1992locking,saad2003iterative,trefethen1997numerical}. In shared clusters or simulation services, adversarial FEM programs can therefore monopolize computational resources, increase operational costs, and delay other users, potentially enabling computational denial-of-service (DoS) attacks~\citep{crosby2003dos,staicu2018freezing,davis2018redos}. Unlike malformed-input attacks, these adversarial programs may remain runnable and complete successfully, making them difficult to distinguish from legitimate but computationally expensive workloads.

This raises a fundamental question about the security of FEM solvers against efficiency attacks:
\begin{center}
\begin{tcolorbox}[colback=blue!6,
colframe=black,
width=\columnwidth,
arc=1mm,
boxrule=0.8pt]

\textit{Can an adversary systematically generate syntactically valid and runnable FEM programs that induce computational resource exhaustion in FEM solvers?}
\end{tcolorbox}
\end{center}

Existing attack techniques are poorly matched to this threat model for three reasons: (1) \textit{heterogeneous attack surfaces}: FEM solvers use different input languages, APIs, data structures, and configuration formats, making it difficult to define a unified attack space across solver stacks; (2) \textit{compileable and runnable constraints}: adversarial changes to geometry, meshes, materials, boundary conditions, or loads must preserve syntactic validity and satisfy runnable and numerical constraints, while arbitrary modifications can easily yield invalid or meaningless programs; and (3) \textit{limited search guidance under black-box access}: unlike attacks on differentiable neural networks, FEM solvers provide no gradients or fine-grained optimization signals. The attacker observes only execution outcomes over a vast, structured program space, making it difficult to identify inputs that induce poor conditioning, slow convergence, or other expensive solver behavior.

To address these challenges, our key insight is to combine \emph{solver-independent abstraction}, \emph{FEM domain knowledge}, and \emph{LLM-guided black-box search}. Based on this insight, we introduce \tool, a domain-guided LLM-agent framework for generating runnable adversarial FEM programs that amplify victim-side computation. First, \tool maps heterogeneous native FEM specifications into a solver-independent intermediate representation (IR) that exposes editable materials, boundary conditions, loads and sources, geometry, and mesh perturbations while protecting solver configurations. Second, it incorporates FEM domain knowledge and deterministic compileablity and runnablity checks to constrain modifications and ensure valid executable programs. Third, an LLM agent reasons over these semantic components and iteratively refines candidate programs using solver execution feedback, steering the search toward increasingly expensive behavior without requiring gradient information.

We evaluate \tool across three FEM stacks, 17 problem classes, and 1,068 configurations against five generic black-box baselines under the same solver-query budget. \tool finds runnable adversarial FEM programs that increase solver work by up to \emph{\(5{,}752.72\times\)}, from 43 to 247,367 iterations. It outperforms generic black-box methods in the unrestricted setting and retains a smaller advantage over evolutionary search under a matched bounded action space. Cross-solver replay further shows that amplification can stem from both problem-intrinsic numerical difficulty and implementation- or discretization-specific sensitivity. Overall, these results expose an underexplored efficiency-attack surface in FEM software.

Our contributions are summarized as follows:

\begin{itemize}

\item \textbf{Problem Novelty.} We identify and systematically study \emph{efficiency attacks} against FEM solvers, in which syntactically valid and runnable FEM programs induce excessive victim-side computation and potentially enable computational DoS.

\item \textbf{Technique Novelty.} We propose \tool, a domain-knowledge-guided LLM-agent framework that combines a solver-independent FEM IR, validity-aware adversarial transformations, and feedback-guided black-box search to generate computationally expensive FEM programs across heterogeneous solver stacks.

\item \textbf{Empirical Evaluation.} Across three FEM stacks, 17 problem classes, and 1,068 configurations, \tool exposes substantial computational amplification, reaching up to \emph{\(5{,}752.72\times\)} solver iterations, and outperforms generic black-box search methods under both unrestricted and controlled comparison settings.

\end{itemize}

To the best of our knowledge, this is the first large-scale study of LLM-agent-guided efficiency attacks against FEM solvers through runnable adversarial FEM program generation.

\section{Background \& Related Work}
\label{sec:related}

\subsection{Finite Element Solvers and Their Numerical Cost}

\begin{figure}[t]
\centering
\IfFileExists{figures/fig_fem_overview.pdf}{
  \includegraphics[width=0.92\linewidth]{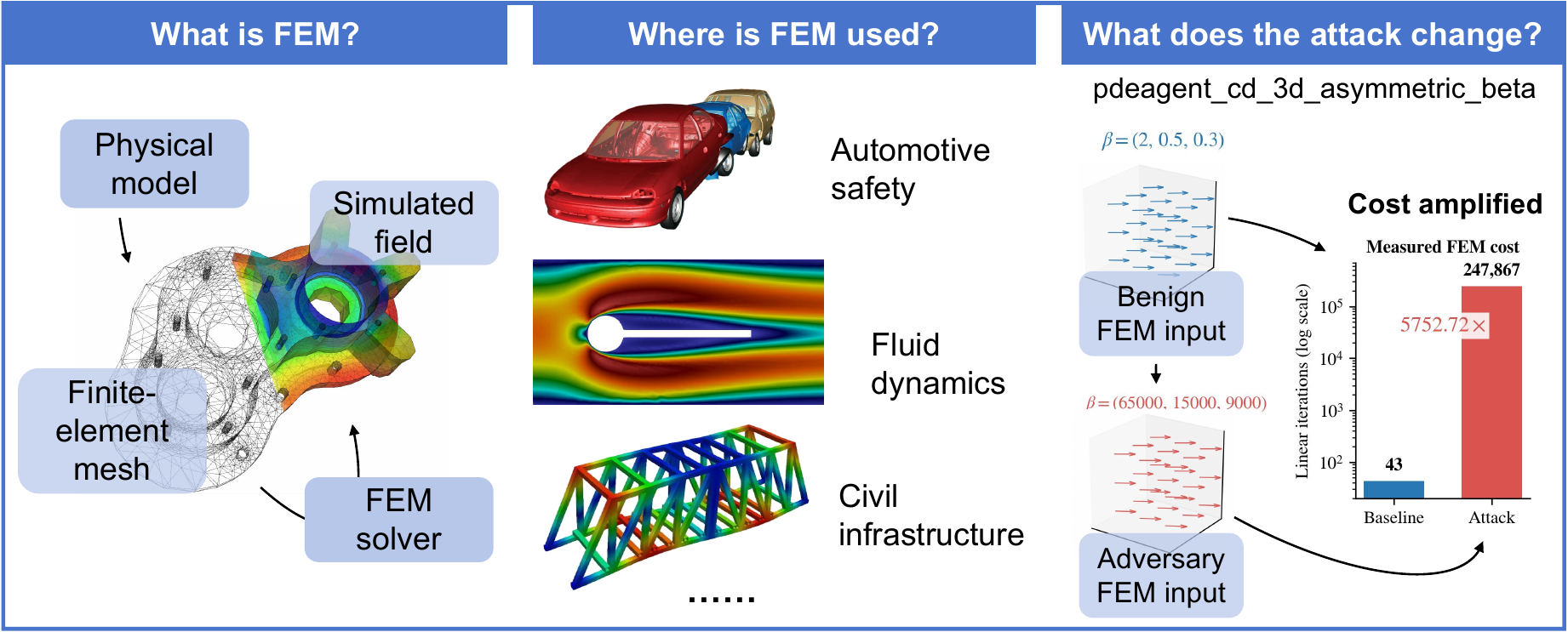}
}{
  \fbox{\parbox[c][0.06\textheight][c]{0.88\linewidth}{
    \centering
    Physical model $\rightarrow$ mesh and discretization $\rightarrow$ sparse solve $\rightarrow$ engineering prediction\\[-1pt]
    \footnotesize Automotive, infrastructure, fluid, and multiphysics applications
  }}
}
\caption{Finite-element workflow and representative applications. Images are adapted from \cite{fem_pump,fem_crash,fem_bridge,fem_fluid}.}
\label{fig:fem-overview}
\end{figure}

FEM maps geometry, materials, boundary conditions, and loads through meshing and weak-form assembly to a sparse system $A(x)u=f(x)$ and an engineering prediction~\citep{hughes2000fem,strang2008analysis}. Figure~\ref{fig:fem-overview} locates our attack surface in this workflow. Physical inputs can amplify solver cost between specification and prediction even at similar mesh sizes: direct cost depends on fill, whereas iterative cost depends on the spectrum and preconditioner~\citep{george1981sparse,davis2006direct,saad2003iterative,trefethen1997numerical}. Near incompressibility, coefficient contrast, and mesh distortion can pass recognized checks yet remain numerically difficult~\citep{babuska1992locking}; gate acceptance implies neither tractability nor application realism. Appendix~\ref{app:related-work} expands the numerical mechanisms and distinguishes well-posedness from conditioning.

\subsection{Efficiency Degradation and Algorithmic Complexity Attacks}

Algorithmic-complexity attacks induce disproportionate work through legitimate inputs, from hash collisions and ReDoS~\citep{crosby2003dos,staicu2018freezing,davis2018redos} to prolonged learned inference~\citep{shumailov2021sponge,hong2021sloth,chen2022nmtsloth,feng2024llmeffichecker, chen2023dark, chen2022nicgslowdown, dong2025engorgio,shapira2023phantom}. Unlike accuracy attacks on scientific surrogates~\citep{adesoji2022fno}, we target deterministic FEM work while solver controls remain protected. Appendix~\ref{app:related-work} gives a detailed comparison with classical algorithmic-complexity and learned-inference slowdown attacks.

\subsection{Validity-Constrained Agentic Search}

Code agents use execution feedback, while FunSearch, AlphaEvolve, and related work place language models inside externally evaluated optimization loops~\citep{chen2025benchmarking, jimenez2024swebench,yang2024sweagent,wang2025openhands,romeraparedes2024funsearch,novikov2025alphaevolve,yang2024opro,liu2024eoh}. They complement vector-based derivative-free methods such as CMA-ES~\citep{hansen2001cmaes}. Here the search object is a heterogeneous FEM specification, every evaluation invokes a solver, and checks and accounting remain deterministic. We test whether semantic access improves evaluator-query efficiency over five black-box baselines; Section~\ref{sec:approach} details the solver-blind agent design, and Appendix~\ref{app:related-work} provides a fuller positioning against code agents and LLM-guided optimization.

\section{Approach}
\label{sec:approach}

\subsection{Threat Model}

\begin{figure}[t]
\centering
\includegraphics[width=0.95\linewidth,height=0.20\textheight,keepaspectratio]{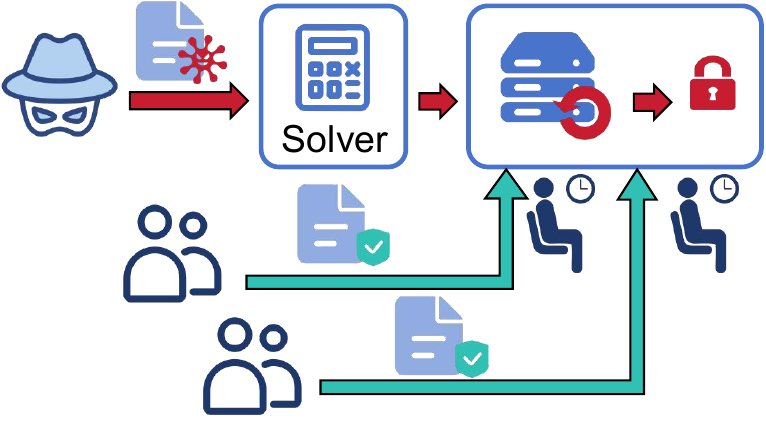}
\caption{Threat model.}
\label{fig:threat-model}
\end{figure}

\paragraph{Adversarial objective} The attacker acts as an ordinary user of a shared FEM service and submits a valid input designed to take substantially longer to solve than a benign one. As illustrated in Figure~\ref{fig:threat-model}, such an input could hold a worker until completion or termination, reducing cluster capacity and delaying benign jobs when deployment controls do not contain its cost. Our experiments establish controlled work amplification rather than production-service resource exhaustion.

\paragraph{Capabilities and knowledge} The attacker has black-box access to the normal submission interface. Starting from a valid seed, it may edit exposed materials, conditions, geometry, and supported mesh-node fields, but not solver methods, tolerances, limits, or mesh resolution (Appendix Table~\ref{tab:ir-surfaces}). It observes past edits, gate outcomes, cost ratios, and service-visible readings such as iterations, wall time, and degrees of freedom. It does not know the query budget and cannot inspect assembled operators, spectra, factorizations, preconditioner state, or hidden solver configuration.

\setlength{\intextsep}{4pt}

\subsection{Problem Formulation}

Let $S$ be an FEM solver with fixed configuration, $x_0$ a valid seed, $\mathcal{R}(x_0)$ its supported edit space, and $C(S,x)$ the cost measure fixed before search. CloakFEM solves
\begin{equation}
\begin{aligned}
x^\star
&=\arg\max_{x\in\mathcal{R}(x_0)\cap\mathcal{X}_{\mathrm{gate}}}
\rho(x),
&\rho(x)&=\frac{C(S,x)}{C(S,x_0)}.
\end{aligned}
\label{eq:objective}
\end{equation}
Here, $\rho$ measures resource amplification relative to the seed under the same solver and cost field. We use linear or nonlinear iteration count when available and wall time otherwise.

The feasible set requires an edit to be representable in the IR, contain finite values, preserve valid geometry, satisfy recognized constitutive and problem-specific runnable constraints, and complete with a converged solve. These checks ensure that edited seeds remain compileable, runnable inputs while excluding malformed or unfinished jobs from successful attacks. Appendix~\ref{app:validity-derivation} gives the gate definition and the elasticity result.

\subsection{CloakFEM Overview}
\label{sec:cloakfem-overview}

\begin{figure}[t]
\centering
\includegraphics[width=0.88\linewidth]{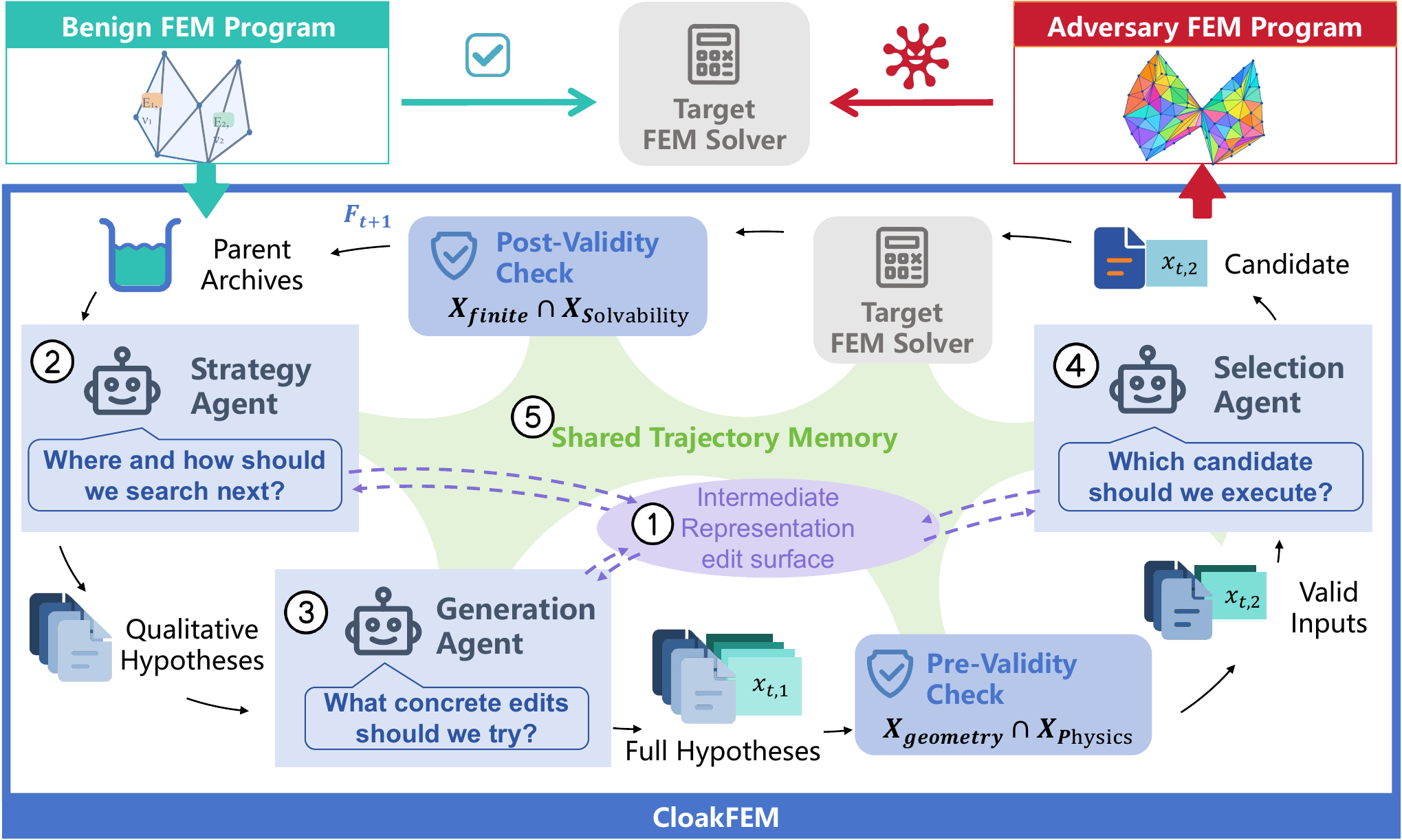}
\caption{CloakFEM Overview.}
\label{fig:cloakfem-overview}
\end{figure}

Figure~\ref{fig:cloakfem-overview} summarizes one CloakFEM iteration:
\textbf{(1) IR interface:} a native FEM input is mapped to the shared IR, and an archived state is selected as the parent;
\textbf{(2) Strategy:} the Strategy Agent chooses physical edit targets, operations, and qualitative directions;
\textbf{(3) Generation:} the Generation Agent instantiates these moves as concrete IR patches, while deterministic pre-validation removes invalid candidates;
\textbf{(4) Selection:} the Selection Agent compares the surviving patches and commits one for execution; and
\textbf{(5) Trajectory update:} deterministic components reconstruct the native input, run the solver, validate the result, and record the outcome in memory and the archive.
The agents operate only on the IR and observed history; native write-back, execution, validation, query accounting, and final ranking remain outside the models. After $B$ queries, CloakFEM returns the accepted input with the largest observed cost ratio.

\subsection{Detailed Design}

\paragraph{IR interface} Native FEM formats express the same physical concepts through different syntax. CloakFEM maps each seed to a solver-independent IR containing its physics label, materials, conditions, supported geometry, and mesh-node edits, while omitting solver configuration. The agents exchange patches over this object rather than native files. Each atomic edit $(o,p,v)$ applies a \texttt{set}, \texttt{add}, or \texttt{delete} operation $o$ to a typed path $p$, with value $v$ when required; a patch may coordinate several such edits. Typed paths make each target and resulting diff explicit, and source-specific writers reconstruct the native input after pre-execution checks. The same representation is used throughout proposal, validation, trajectory memory, and write-back, giving different FEM backends a common interface. Appendix Table~\ref{tab:ir-surfaces} lists the edit surfaces, and Appendix~\ref{app:ir-example} gives a complete example.

\paragraph{Strategy Agent} From the IR, archive, and prior outcomes, it proposes up to three prioritized moves specifying targets, operations, directions, and optional parents, but no numeric magnitudes. This separates \emph{where} to search from \emph{how far} to edit. Multiple moves allow distinct compileable hypotheses to be considered in the same iteration.It uses trajectory history to avoid failed directions and branch from promising archived states.

\paragraph{Generation Agent} It instantiates each move at several magnitudes as concrete, potentially multi-field IR patches. Deterministic checks reject malformed, unchanged, unsupported, or immediately invalid candidates before solver execution; rejection reasons may guide a retry without consuming a solver query. This magnitude ladder balances exploration and exploitation.

\paragraph{Selection Agent} From the surviving candidates and their concrete diffs, it predicts cost ratios and selects one move--magnitude pair for execution. It may override earlier preferences using accumulated evidence, enabling broader exploration while spending one solver query. This focuses solver computation on the candidate judged most promising at that iteration. Predictions allocate queries; solver execution and the gate determine outcomes. Appendix~\ref{app:agent-prompts} provides all three agent prompts.

\paragraph{Trajectory memory and archive} A bounded summary of attempts, target-level outcomes, solver feedback, and provenance supports credit assignment and avoids repeated ineffective edits. The archive retains every gate-accepted state, allowing later moves to branch from diverse valid parents. Together, they balance refinement of strong candidates with exploration of alternative search paths. Implementation details appear in Appendix~\ref{app:search-details}.

\section{Evaluation}
\label{sec:evaluation}

\IfFileExists{experimental_setup.tex}
  {\subsection{Experimental Setup}
\label{sec:setup}

\paragraph{Datasets} We collect 1,746 public FEM specifications from the Elmer regression suite, PDE-Agent-Bench, and NAFEMS~\citep{elmer2013,hang2026pdeagentbench,nafems1990benchmarks}. Baseline replay and edit-surface checks produce a fixed 1,068-seed evaluation cohort containing 618 Elmer, 444 PDE-Agent, and 6 NAFEMS cases. Appendix~\ref{app:dataset-details} reports provenance, representations, physics coverage, target-solver mappings, and exclusions.

\paragraph{Target solvers} Elmer cases run in native ElmerSolver~\citep{elmer2013}; PDE-Agent cases use DOLFINx 0.9, PETSc 3.24, and Gmsh 4.15.2~\citep{baratta2023dolfinx,balay2019petsc,geuzaine2009gmsh}; and supported NAFEMS cases use scikit-fem 12.0.2 or the DOLFINx scalar bridge~\citep{gustafsson2020scikitfem}.

\paragraph{Baselines} We compare CloakFEM against uniform random search, an evolutionary strategy, CMA-ES~\citep{hansen2001cmaes}, and the Nevergrad OnePlusOne and CMA implementations~\citep{rapin2018nevergrad}. All methods share the manifest, IR, writers, target solvers, compileablity and runnability policy, and nominal $B=40$ solver-query cap .

\paragraph{Implementation}The primary RQ1 campaigns use \texttt{deepseek-flash} and \texttt{gpt-5.6-luna} at temperature $0.7$, with the same prompt templates, feedback fields, query budget, requested candidate width $k=2$, and at most three Strategy moves per iteration. Framework-controlled random choices and stochastic baselines use replicate seed 0, while the model APIs use their standard sampling behavior without an explicit seed. ElmerGrid, individual solver executions, and seed trajectories are capped at 60, 300, and 1,800 seconds, respectively; incomplete trajectories retain their checkpointed best gate-accepted result. Runs are distributed across independent process shards, with numerical-library threads fixed to one, on two AMD EPYC 7352 processors (96 hardware threads) with 503\,GB RAM; target solvers do not use GPUs. Each record stores model-request outcomes, input, cache, and output tokens, and end-to-end trajectory wall time covering model, solver, validation, and orchestration. Appendix~\ref{app:dataset-details} gives additional execution and logging details.

}
  {}

\subsection{RQ1: How Much Cost Amplification Can \tool Find?}
\label{sec:rq1}

\subsubsection{RQ1.1: Prevalence and Severity}

\begin{table*}[t]
\centering
\footnotesize
\setlength{\tabcolsep}{3.5pt}
\caption{Count-based amplification occurs across all three datasets.}
\label{tab:rq1-existence}
\begin{tabular*}{0.80\textwidth}{@{\extracolsep{\fill}}lrrrrrr@{}}
\hline
\multicolumn{1}{c}{}
& \multicolumn{3}{c}{CloakFEM (DeepSeek)}
& \multicolumn{3}{c}{CloakFEM (GPT)} \\
\cline{2-7}
$\rho^\star$ & Elmer & PDE-Agent & NAFEMS
& Elmer & PDE-Agent & NAFEMS \\
\hline
Median & 3.01 & 1.54 & 8.72 & 5.00 & 1.93 & 11.85 \\
Arithmetic mean & 24.17 & 90.30 & 17.88 & 39.59 & 131.34 & 22.77 \\
Geometric mean & 4.79 & 3.32 & 7.52 & 7.24 & 4.92 & 8.37 \\
P75 & 12.08 & 6.38 & 17.22 & 17.18 & 11.03 & 36.47 \\
P90 & 42.33 & 24.00 & 43.36 & 97.04 & 171.21 & 55.30 \\
Max & 714.29 & 5715.53 & 67.93 & 714.29 & 5752.72 & 68.24 \\
$\Pr(\rho^\star>1.5)$ & 68.9\% & 52.3\% & 83.3\% & 79.1\% & 61.3\% & 66.7\% \\
$\Pr(\rho^\star>3)$ & 50.2\% & 32.0\% & 66.7\% & 63.0\% & 39.6\% & 66.7\% \\
$n$ & 235 & 444 & 6 & 235 & 444 & 6 \\
\hline
\end{tabular*}
\end{table*}

\paragraph{Metric and evaluation cohort} For a seed $x_0$ and a gate-accepted candidate $x$, amplification is $\rho(x)=C(S,x)/C(S,x_0)$, and $\rho^\star$ is the largest gate-accepted value observed under the nominal cap of $B=40$ solver executions. Linear or nonlinear iteration counts are preferred as deterministic work metrics; wall time is used only when no usable count is available, with one cost field locked per seed before search. The fixed 1,068-seed cohort therefore contains 685 count-based and 383 wall-time cases. Table~\ref{tab:rq1-existence} reports only the count-based cases, with each seed contributing one best finite ratio; Appendix~\ref{app:dataset-details} gives the metric-assignment rule.

\paragraph{Amplification prevalence} Table~\ref{tab:rq1-existence} and Figure~\ref{fig:rq1-combined}(a) show a strongly long-tailed distribution across all three datasets. CloakFEM (DeepSeek) reaches median and P90 amplification of $1.89\times$ and $31.41\times$, while CloakFEM (GPT) reaches $2.60\times$ and $127.01\times$. Both configurations find high-cost inputs in every dataset source, with the strongest upper tail on PDE-Agent.

\paragraph{Upper-tail robustness} The largest GPT case raises linear work from 43 to 247,367 iterations ($5{,}752.72\times$), so the maximum is not driven by a one-iteration baseline. Requiring a baseline count of at least three still yields P90 $138.33\times$ and maximum $5{,}752.72\times$ for GPT. The separate wall-time stratum also exhibits a persistent upper tail under repeated measurement. Appendix~\ref{app:rq1-complete} reports the wall-time confirmation, denominator sensitivity, dataset-wise variation, incomplete-run accounting, and representative cases.

\subsubsection{RQ1.2: Search Effectiveness}

\paragraph{Comparison with generic search} Under the common 40-query cap on the matched count cohort, CloakFEM (GPT) reaches median amplification $2.62\times$, P90 $156.0\times$, and a 67.9\% rate above $1.5\times$; evolutionary search reaches $1.14\times$, $6.47\times$, and 29.9\%, respectively. CloakFEM also reaches $\rho>2$ after a median of four queries, compared with twelve for evolutionary search. Appendix Table~\ref{tab:rq1-search} reports all seven methods, the separate wall-time stratum, valid-candidate rates, and incomplete trajectories; Appendix Table~\ref{tab:common-query-prefix} confirms the ranking at matched 10- and 20-query prefixes.

\paragraph{Strict-relative comparison} Table~\ref{tab:rq1-restricted-actions} gives CloakFEM (GPT) and evolutionary search the same bounded action space on all 444 PDE-Agent seeds: nonzero numeric values retain their sign and stay within $0.5$--$2\times$ their seed magnitude, zeros and structure remain fixed, $\nu\leq0.49$, and cumulative node displacement is at most $0.05h$. On the 411 finite pairs, the geometric mean of $\rho_{\text{CloakFEM}}/\rho_{\text{evolutionary}}$ is 1.147 (95\% CI [1.114, 1.183]), while CloakFEM's success rate above $1.5\times$ is 9.5 percentage points higher (95\% CI [$+6.6$, $+12.7$] pp). Both paired tests reject the null with $p<10^{-9}$. Appendix Tables~\ref{tab:strict-evolutionary-repeats} and~\ref{tab:boundary-prior-subset} additionally report three framework seeds for the evolutionary reference and a deterministic non-LLM boundary-prior baseline.

\begin{table*}[t]
\centering
\footnotesize
\caption{CloakFEM retains an advantage under strict-relative edits.}
\label{tab:rq1-restricted-actions}
\begin{minipage}[t]{0.45\linewidth}
\centering
\textit{(a) Method-level outcomes}\\[2pt]
\setlength{\tabcolsep}{1.5pt}
\begin{tabular}{@{}lrrrr@{}}
\hline
Method & Finite & Median & P90 & $>1.5$ \\
\hline
Evolutionary & 439/444 & 1.017 & 1.676 & 14.8\% \\
CloakFEM (GPT) & 412/444 & 1.038 & 3.057 & 25.2\% \\
\hline
\end{tabular}
\end{minipage}
\hfill
\begin{minipage}[t]{0.52\linewidth}
\centering
\textit{(b) Paired advantage of CloakFEM (GPT), $n=411$}\\[2pt]
\setlength{\tabcolsep}{3.0pt}
\begin{tabular}{@{}lcc@{}}
\hline
Outcome & Effect [95\% CI] & $p$ \\
\hline
\shortstack[l]{Geometric mean of\\paired ratios}
& \shortstack{1.147\\{[1.114, 1.183]}}
& $4.17{\times}10^{-22}$ \\
\noalign{\vskip 1.2pt}
\cline{1-3}
\noalign{\vskip 1.2pt}
\shortstack[l]{Success-rate\\difference}
& \shortstack{$+9.5$ pp\\{[$+6.6$, $+12.7$] pp}}
& $2.15{\times}10^{-10}$ \\
\hline
\end{tabular}
\end{minipage}
\end{table*}

\begin{figure*}[!t]
\centering
\includegraphics[width=0.8\textwidth]{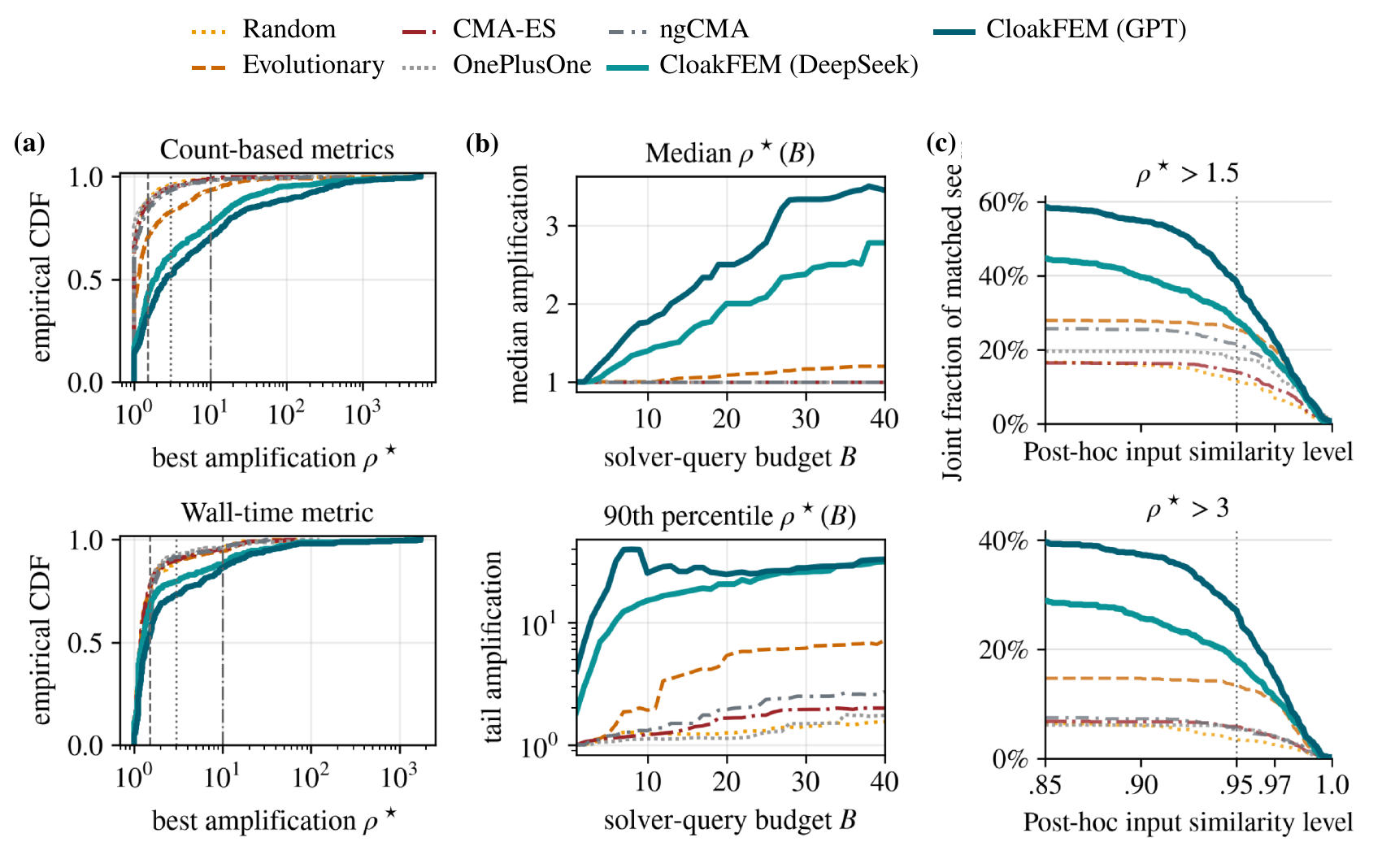}
\caption{Observed amplification, query-prefix progress, and input similarity.}
\label{fig:rq1-combined}
\end{figure*}

\paragraph{Similarity and budget efficiency} At post-hoc overall input similarity $S_{\mathrm{overall}}\geq0.95$, 38.3\% of CloakFEM (GPT) winners exceed $\rho^\star>1.5$ and 27.1\% exceed $3\times$, compared with 25.7\% and 13.5\% for evolutionary search. Figure~\ref{fig:rq1-combined}(b--c) shows that CloakFEM builds its advantage early in the query budget and preserves it across most similarity thresholds. Similarity is descriptive rather than a search constraint; Appendix~\ref{app:input-similarity} defines its structured-IR components, and Appendix~\ref{app:rq1-complete} provides the complete search, cost, and similarity tables.

\FloatBarrier
\subsection{RQ2: Does the Degradation Transfer Across Solvers?}
\label{sec:rq2}

\paragraph{Transfer protocol} We replay each original seed and its exact saved CloakFEM (DeepSeek) winner in two independent FEM implementations without additional search, parameter adjustment, or model calls. The 100 cases span elasticity, Poisson, convection--diffusion, Stokes, and Navier--Stokes. Appendix Table~\ref{tab:rq2-transfer} reports median work amplification within each solver using CG, GMRES, MINRES, or cumulative Newton iterations. Transfer requires both independent implementations to exceed $1.5\times$; magnitude reproduction additionally requires both ratios to remain within a factor of two of the native result.

\paragraph{Problem-intrinsic versus implementation-specific degradation} Elasticity transfers consistently: 42 of 45 attacks reproduce in both canonical implementations and production ElmerSolver, including all 41 native attacks above $2\times$. Median amplification remains close across the source implementation, scikit-fem, DOLFINx, and Elmer ($10.99\times$, $10.25\times$, $10.25\times$, and $10.69\times$). Every edit that moves Poisson's ratio toward the incompressible limit reproduces, identifying volumetric locking as a problem-intrinsic mechanism. Poisson shows a complementary pattern: seven of ten attacks reproduce across the two canonical assemblies, while production Elmer retains the strongest effect for a smaller subset, indicating discretization-dependent sensitivity.

\begin{figure*}[!t]
\centering
\includegraphics[width=0.85\textwidth]{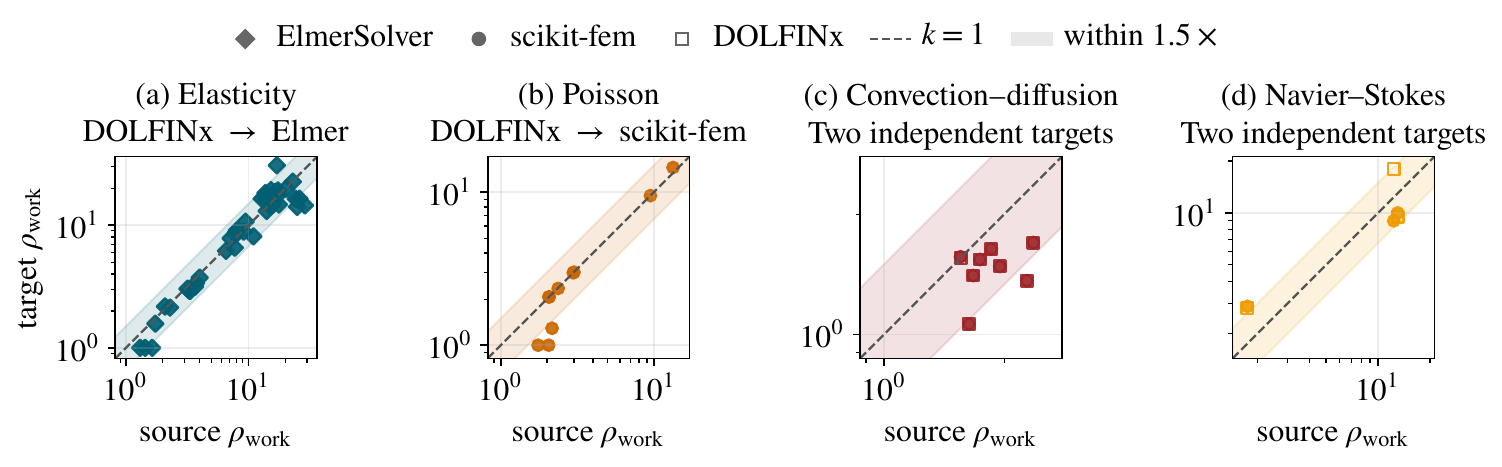}
\caption{Cross-solver transfer varies by physical system.}
\label{fig:rq2-transfer}
\end{figure*}

\paragraph{Transfer across operator classes} The result extends beyond symmetric positive-definite systems. Four of eight convection--diffusion attacks preserve native-scale amplification under GMRES, and 17 of 34 Stokes attacks remain above $1.5\times$ under MINRES on both assemblies. All three Navier--Stokes viscosity attacks transfer under independent Taylor--Hood/Newton implementations, with median amplification $9.00\times$ and $9.50\times$ versus $12.33\times$ natively. Figure~\ref{fig:rq2-transfer} visualizes the cross-solver agreement; Appendix~\ref{app:rq2-complete} provides the complete replay results, and Appendix~\ref{app:rq2-mechanisms} details the numerical mechanisms.

\FloatBarrier
\subsection{RQ3: Which Search Choices Matter?}
\label{sec:rq3}

\subsubsection{RQ3.1: Component Ablation}

\paragraph{Ablation design} We evaluate fresh CloakFEM (GPT) trajectories on PDE-Agent seeds while fixing the IR edit surface, validity gate, unaffected prompts, and 40-query cap. The complete Strategy--Generation--Selection pipeline and all controls request candidates per move except the explicit $k=1$ control. The controls remove or replace the FEM primer, solver feedback, candidate width, learned selection, Strategy Agent scheduling, parent selection, or Generation Agent knowledge.

\begin{figure*}[!t]
\centering
\includegraphics[width=0.80\textwidth]{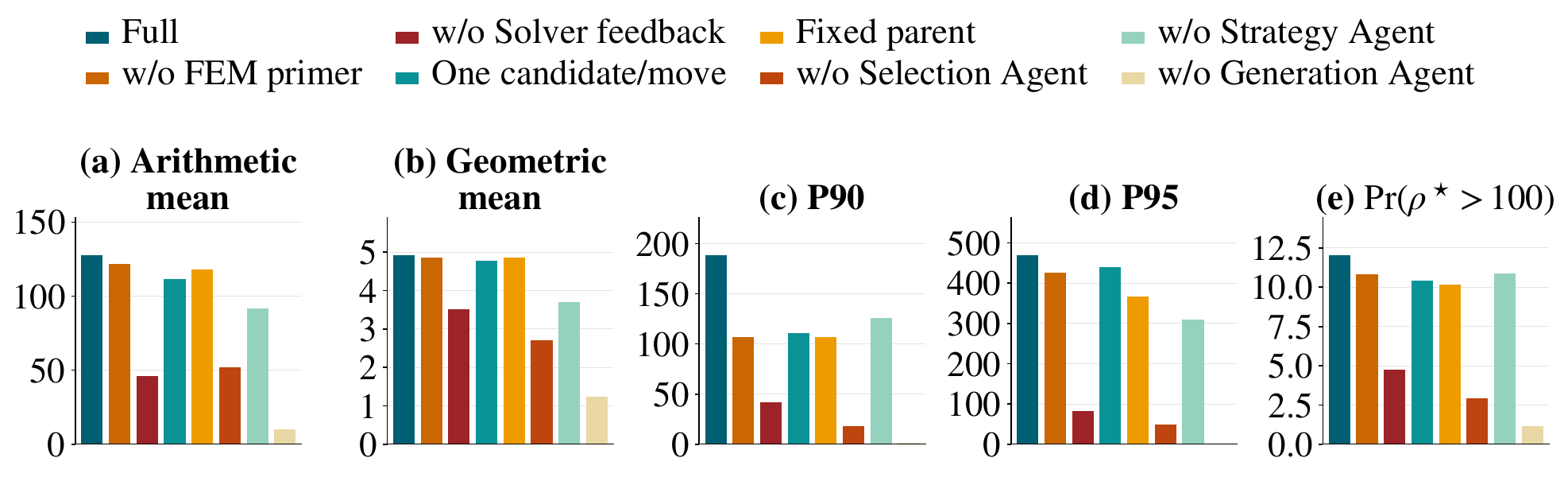}
\caption{Ablations relative to the complete Full (IR) pipeline.}
\label{fig:rq3-ablation}
\end{figure*}

\paragraph{Most influential components} The zero-knowledge Generation control produces the largest decline in Figure~\ref{fig:rq3-ablation}: geometric-mean amplification falls from 4.91 to 1.24, P90 from 188.13 to 1.46, and $\Pr(\rho^\star>100)$ from 12.0\% to 1.1\%. Removing solver feedback or learned selection also lowers the geometric mean to 3.52 and 2.70 and P90 to 42.09 and 17.66, respectively. The higher valid-candidate rate without learned selection (85.1\% versus 74.9\%) further shows that validity alone does not identify high-cost candidates. Appendix~\ref{app:rq3-complete} reports all arms, coverage, and effect summaries.

\subsubsection{RQ3.2: Budget Sensitivity}
Figure~\ref{fig:rq1-combined}(b) traces the fixed 680-seed matched count cohort through the standard budget $B=40$: CloakFEM separates from generic search early and carries its best gate-accepted candidate forward at each query prefix, complementing the median query-to-$2\times$ result in Appendix Table~\ref{tab:rq1-search}. In a separate run of the complete CloakFEM (GPT) pipeline on all 444 PDE-Agent seeds through $B=100$, the fixed-cohort median best-so-far amplification rises from 2.05 at $B=40$ to 2.42 at $B=100$, a 17.7\% gain, while P90 rises from 125.89 to 150.93. The improvement is strongest before $B=80$, after which the upper tail plateaus. Appendix Figure~\ref{fig:rq3-budget-sensitivity} and Table~\ref{tab:rq3-budget-sensitivity} provide the complete curves, intermediate budgets, and coverage.

\FloatBarrier
\subsection{RQ4: What Mitigation Tradeoffs Appear in the Implemented Cells?}
\label{sec:rq4}

\subsubsection{RQ4.1: Runtime Limits}

\begin{wrapfigure}[12]{r}{0.40\textwidth}
\centering
\includegraphics[width=0.96\linewidth]{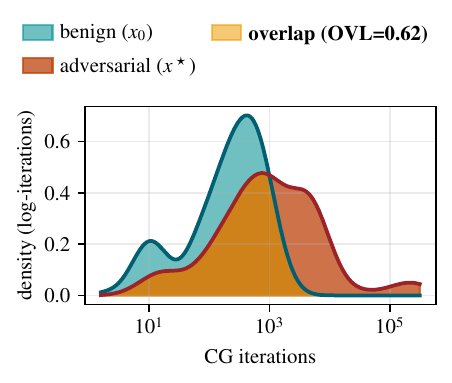}
\captionsetup{skip=2pt}
\caption{Benign and adversarial iteration distributions overlap substantially.}
\label{fig:rq4-timeout}
\end{wrapfigure}

Figure~\ref{fig:rq4-timeout} compares benign and adversarial iteration counts across 171 CG seeds. Their distributions have an overlap coefficient of 0.62, showing that a single iteration threshold has limited discriminative power. At a 1\% false-positive rate, the threshold detects 34.5\% of adversarial inputs; increasing detection therefore requires rejecting a larger fraction of legitimate expensive jobs, so a single runtime threshold cannot separate adversarial from benign load. Appendix Figure~\ref{fig:rq4-timeout-sweep} reports the complete detection curve across false-positive allowances.

\subsubsection{RQ4.2: Solver-Side Defenses}

\paragraph{Defense evaluation} The strict-pair matrix applies each solver configuration to all applicable benign--edited pairs and evaluates three properties together: benign runtime overhead, residual edited-to-benign amplification, and paired completion. Across 535 replays, 526 complete as pairs. Appendix Table~\ref{tab:rq42-strict-followup} reports the complete seven-system matrix; no configuration achieves low residual amplification, low benign overhead, and full completion at once, and Appendix Figure~\ref{fig:rq42-system} visualizes this tradeoff.

\paragraph{Elasticity defenses} On 45 elasticity seeds, Jacobi is effectively unchanged from CG. AMG completes all pairs but costs $5.01\times$ on benign inputs and leaves $2.12\times$ time amplification. GMRES--ILU lowers residual amplification to $1.26\times$ in time at $10.31\times$ benign overhead, while direct LU completes all pairs with a $0.98\times$ edited-to-benign time ratio and $2.00\times$ benign overhead at the tested scale. This clean time behavior is scale-bounded, not general: the direct factors store a median $11\times$ (P90 $16.4\times$) the original nonzeros, and since sparse-direct fill grows superlinearly, the same defense becomes prohibitive as the problem size grows.

\paragraph{Nonlinear mitigation} Viscosity continuation provides the clearest nonlinear result: on 22 Navier--Stokes seeds it reduces the cumulative-Newton ratio from $1.33\times$ to $1.03\times$ with 22/22 paired completion, at $2.38\times$ benign overhead. Appendix~\ref{app:rq4-complete} retains the control systems, scale analysis, and complete replay accounting.

\FloatBarrier

\section{Conclusion}
\label{sec:conclusion}

CloakFEM exposes input-induced solver cost as a resource-exhaustion vulnerability surface in FEM software. Its solver-independent IR and agentic search discover gate-accepted inputs while protecting solver controls, with the strongest count-based case rising from 43 to 247,367 iterations and substantial amplification across the long-tailed cohort. Strict-relative and matched-prefix controls support the search advantage, while cross-solver replay separates problem-intrinsic difficulty from implementation-specific sensitivity. Together, these results establish CloakFEM as a vulnerability-oriented red-teaming framework for identifying costly FEM inputs whose deployment risk depends on concrete access controls, quotas, and service guarantees.

\section*{Statement}

\subsection*{AI use statement}
Generative-AI assistants were used for English-language editing and for assistance with code, plotting, and debugging. They were not treated as sources of scientific evidence. The authors designed the methodology, selected and curated the data, executed the experiments, verified the reported results, reviewed all generated text and code, and take responsibility for the final manuscript. The language-model calls made by CloakFEM are part of the evaluated research method rather than manuscript assistance; their configurations, prompts, execution feedback, and usage are reported or logged as experimental artifacts.

\subsection*{Ethics statement}
This work studies input-induced computational resource exhaustion in FEM solvers to expose a robustness and security risk that conventional correctness- and crash-oriented testing can overlook, not to facilitate disruption of deployed services. All experiments were executed in a controlled research environment on public FEM benchmarks; no production service, user account, human subject, or personal data was involved. CloakFEM edits exposed FEM inputs but cannot alter protected solver methods, tolerances, stopping limits, or resource controls. We report acceptance checks, cross-solver attribution, limitations on deployment-level vulnerability claims, and solver-side mitigations in Section~\ref{sec:rq4}. These safeguards and disclosures are intended to support defensive stress testing and resource-governance design.

\subsection*{Reproducibility statement}
Section~\ref{sec:approach} specifies the threat model, objective, IR, and agent pipeline, while Section~\ref{sec:setup} reports the datasets, solver versions, baselines, budgets, timeouts, and hardware. Appendices~\ref{app:search-details}, \ref{app:validity-derivation}, and \ref{app:dataset-details} provide the exact search policies, acceptance checks, corpus provenance, and exclusions; Appendix~\ref{app:reproducibility} documents record identity, execution boundaries, randomness, and generated artifacts. Framework-controlled sampling and black-box baselines use replicate seed 0, numerical-library threads are fixed to one, and the language-model APIs run at temperature 0.7 without an explicit sampling seed. Consequently, framework decisions are replayable from the logged state, but end-to-end language-model trajectories are not bitwise deterministic. Unless stated otherwise, each seed--method pair has one trajectory. The artifact retains prompts, configurations, immutable seed identifiers, candidate and validity logs, query accounting, model usage, and the scripts and intermediate data used to regenerate the reported analyses.

\bibliographystyle{plainnat}
\bibliography{references}
\clearpage
\appendix
\section*{Appendix Contents}
\begin{small}
\noindent\hyperref[app:complete-evaluation]{\ref*{app:complete-evaluation}\quad Complete Evaluation Results}\dotfill\pageref*{app:complete-evaluation}\par
\noindent\hspace*{1.5em}\hyperref[app:rq1-complete]{\ref*{app:rq1-complete}\quad Complete RQ1 Results}\dotfill\pageref*{app:rq1-complete}\par
\noindent\hspace*{1.5em}\hyperref[app:rq2-complete]{\ref*{app:rq2-complete}\quad Complete RQ2 Results}\dotfill\pageref*{app:rq2-complete}\par
\noindent\hspace*{1.5em}\hyperref[app:rq3-complete]{\ref*{app:rq3-complete}\quad Complete RQ3 Results}\dotfill\pageref*{app:rq3-complete}\par
\noindent\hspace*{1.5em}\hyperref[app:rq4-complete]{\ref*{app:rq4-complete}\quad Complete RQ4 Results}\dotfill\pageref*{app:rq4-complete}\par
\noindent\hyperref[app:search-details]{\ref*{app:search-details}\quad Search and IR Implementation Details}\dotfill\pageref*{app:search-details}\par
\noindent\hyperref[app:agent-prompts]{\ref*{app:agent-prompts}\quad Three-Agent Prompts}\dotfill\pageref*{app:agent-prompts}\par
\noindent\hyperref[app:ir-example]{\ref*{app:ir-example}\quad Illustrative IR Edit and Native Write-Back}\dotfill\pageref*{app:ir-example}\par
\noindent\hyperref[app:validity-derivation]{\ref*{app:validity-derivation}\quad Acceptance Gate and Well-Posed PDE Subclasses}\dotfill\pageref*{app:validity-derivation}\par
\noindent\hyperref[app:dataset-details]{\ref*{app:dataset-details}\quad Dataset and Target-Solver Details}\dotfill\pageref*{app:dataset-details}\par
\noindent\hyperref[app:representative-cases]{\ref*{app:representative-cases}\quad Representative FEM Case Visualizations}\dotfill\pageref*{app:representative-cases}\par
\noindent\hyperref[app:input-similarity]{\ref*{app:input-similarity}\quad Input-Similarity Measurement}\dotfill\pageref*{app:input-similarity}\par
\noindent\hyperref[app:rq2-mechanisms]{\ref*{app:rq2-mechanisms}\quad Numerical Mechanisms of Efficiency Degradation}\dotfill\pageref*{app:rq2-mechanisms}\par
\noindent\hyperref[app:budget-sensitivity]{\ref*{app:budget-sensitivity}\quad Extended-Budget Sensitivity}\dotfill\pageref*{app:budget-sensitivity}\par
\noindent\hyperref[app:reproducibility]{\ref*{app:reproducibility}\quad Reproducibility and Reporting Details}\dotfill\pageref*{app:reproducibility}\par
\noindent\hyperref[app:related-work]{\ref*{app:related-work}\quad Extended Background and Related Work}\dotfill\pageref*{app:related-work}\par
\end{small}

\section{Complete Evaluation Results}
\label{app:complete-evaluation}

\subsection{Complete RQ1 Results}
\label{app:rq1-complete}

\paragraph{RQ1.1: Prevalence and severity} For a seed $x_0$ and a gate-accepted candidate $x$, amplification is $\rho(x)=C(S,x)/C(S,x_0)$, and $\rho^\star$ is the largest gate-accepted value observed under the nominal cap of $B=40$ solver executions. Linear or nonlinear iteration counts are preferred as deterministic work metrics; wall time is used otherwise, with one effective cost field locked per seed before search. This produces 685 count-based and 383 wall-time ratios across the 1,068-seed cohort. Appendix~\ref{app:dataset-details} specifies the deterministic metric-fallback rule. Table~\ref{tab:rq1-existence} and Figure~\ref{fig:rq1-combined}(a) reveal a strongly long-tailed count-based distribution across dataset sources and mark the $1.5\times$, $3\times$, and $10\times$ thresholds. Both models produce a finite baseline-relative result for every included seed. On count-based seeds, the median and 90th percentile are respectively $1.89\times$ and $31.41\times$ for CloakFEM (DeepSeek), and $2.60\times$ and $127.01\times$ for CloakFEM (GPT). The largest GPT case is not a one-iteration denominator: work increases from 43 to 247,367 linear iterations, or $5{,}752.72\times$. On wall-time seeds, the corresponding median and 90th percentile are $1.25\times$ and $11.93\times$ for CloakFEM (DeepSeek), and $1.42\times$ and $14.60\times$ for CloakFEM (GPT); GPT's median is below ngCMA's $1.45\times$, while both CloakFEM variants have substantially larger P90 values than all five baselines. Table~\ref{tab:rq1-search} reports this metric stratum separately. Each baseline and candidate is timed once within its trajectory, execution order is not randomized, and the selected winner is not independently remeasured. We therefore treat all wall-time maxima and threshold rates as exploratory rather than confirmed runtime effects. Table~\ref{tab:rq1-existence} contains only count-level summaries and never averages iteration counts with wall time; within each stratum, every seed contributes exactly one ratio from its locked metric.

\paragraph{Coverage and incomplete runs} Table~\ref{tab:rq1-search} reports runs that stop before using all 40 queries, while Figure~\ref{fig:rq1-combined}(a) includes their checkpointed lower bounds. Each solver execution is capped at 300 seconds and each seed trajectory at 1,800 seconds. CloakFEM (DeepSeek) completes all 40 queries on 910 seeds and stops early on 158; CloakFEM (GPT) completes 841 and stops early on 227. An incomplete trajectory retains its checkpointed best gate-accepted result, so its reported $\rho^\star$ is a best-so-far lower bound rather than the outcome of a completed budget. Conversely, a candidate that reaches the 300-second single-solve limit is not counted as a completed result; for a wall-time seed it establishes only the lower bound $\rho\geq 300/C(S,x_0)$. Because high-cost inputs consume the trajectory limit fastest, dropping incomplete runs would preferentially remove difficult seeds. Retaining them avoids that exclusion but does not make realized budgets equal: the primary table compares methods under the same query \emph{cap}, not the same completed query count, wall-clock search time, or total compute. The matched-prefix analysis below equalizes completed queries; explicit survival analysis remains future work.

\paragraph{Matched completed-query prefixes} We additionally compare all seven methods on count-metric seeds for which every method completed the stated prefix. At 10 completed queries, the paired cohort contains 580 seeds: CloakFEM (GPT) has median $\rho^\star=1.762$ versus 1.000 for evolutionary search, with a $+43.4$ percentage-point difference in $\Pr(\rho^\star>1.5)$. At 20 completed queries, the paired cohort contains 508 seeds and the corresponding values are 2.500 versus 1.087 and $+42.3$ percentage points. The paired geometric-mean CloakFEM-to-evolutionary ratios are 3.009 and 2.614 at 10 and 20 queries, respectively. These fixed-prefix results show that the primary advantage is not created by comparing different realized query counts, although they do not equalize total solver or model time. Because the 10- and 20-query columns use different matched cohorts, their percentiles need not be monotone across the two columns.

\begin{table*}[t]
\centering
\footnotesize
\setlength{\tabcolsep}{3.2pt}
\caption{Matched completed-query prefixes preserve the main ranking.}
\label{tab:common-query-prefix}
\begin{tabular*}{\textwidth}{@{\extracolsep{\fill}}lrrrrrr@{}}
\hline
& \multicolumn{3}{c}{$B=10$ ($n=580$)}
& \multicolumn{3}{c}{$B=20$ ($n=508$)} \\
\cline{2-4}\cline{5-7}
Method & Median & P90 & $>1.5$ & Median & P90 & $>1.5$ \\
\hline
Random & 1.000 & 1.228 & 5.9\% & 1.000 & 1.250 & 6.5\% \\
Evolutionary & 1.000 & 1.920 & 12.8\% & 1.087 & 5.358 & 23.8\% \\
CMA-ES & 1.000 & 1.215 & 4.7\% & 1.000 & 1.656 & 10.4\% \\
OnePlusOne & 1.000 & 1.119 & 4.3\% & 1.000 & 1.128 & 5.7\% \\
ngCMA & 1.000 & 1.307 & 7.6\% & 1.000 & 1.938 & 13.4\% \\
CloakFEM (DeepSeek) & 1.393 & 15.121 & 44.7\% & 2.000 & 20.622 & 56.1\% \\
CloakFEM (GPT) & 1.762 & 25.396 & 56.2\% & 2.500 & 24.771 & 66.1\% \\
\hline
\end{tabular*}
\end{table*}

\paragraph{Search-side cost} Across the fixed 1,068-seed cohort, CloakFEM (DeepSeek) uses a median 171 successful model requests, 748,918 input tokens, 80,022 output tokens, and 9.9 minutes of end-to-end trajectory time per seed. CloakFEM (GPT) uses 124 requests, 468,754 input tokens, 48,494 output tokens, and 13.3 minutes per seed. Total recorded trajectory time is 238.1 and 293.6 hours, respectively. Per-seed entries in Table~\ref{tab:search-side-cost} are medians, and time includes model, solver, validation, and orchestration. These values expose substantial proposal-side cost omitted by a solver-query-only budget. The served identifiers \texttt{deepseek-flash} and \texttt{gpt-5.6-luna} came through experiment gateways rather than documented public billable SKUs, so applying a public DeepSeek or OpenAI price would not reproduce the experiment's charge. The logs also do not separate model latency from solver time. We therefore do not claim superiority under equal wall-clock or monetary cost.

\begin{table*}[t]
\centering
\footnotesize
\setlength{\tabcolsep}{4pt}
\caption{Proposal-side cost is substantial and model dependent.}
\label{tab:search-side-cost}
\begin{tabular*}{\textwidth}{@{\extracolsep{\fill}}lrrrrr@{}}
\hline
Method & Calls & Input tokens & Output tokens & Time/seed & Total time \\
\hline
CloakFEM (DeepSeek) & 171 & 748,918 & 80,022 & 9.9 min & 238.1 h \\
CloakFEM (GPT) & 124 & 468,754 & 48,494 & 13.3 min & 293.6 h \\
\hline
\end{tabular*}
\end{table*}

\begin{wrapfigure}[14]{r}{0.40\textwidth}
\centering
\includegraphics[width=0.96\linewidth]{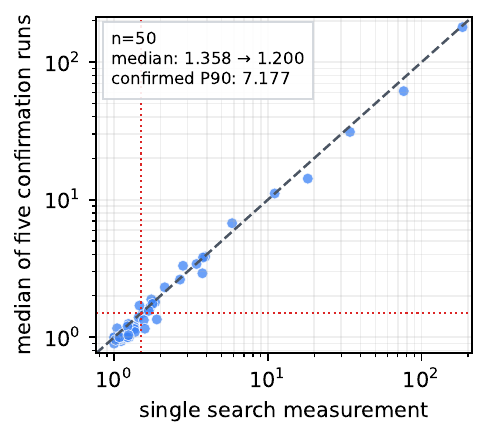}
\caption{Repeated timing attenuates selected wall-time ratios.}
\label{fig:walltime-confirmation}
\end{wrapfigure}

\paragraph{Wall-time confirmation} On a fixed, non-random 50-case Elmer confirmation cohort drawn from saved GPT candidates, we independently execute the baseline and winner five times each, randomizing their order within every repeat and taking the ratio of medians. The selection manifest records the cases but not a probability-sampling design, so this experiment measures repeatability for those cases and is not representative of all 383 wall-time seeds. Figure~\ref{fig:walltime-confirmation} compares the original single search measurement with the ratio of five-run medians. The original median is 1.358; the confirmed median is 1.200 (seed-bootstrap 95\% CI [1.051, 1.576]), with P90 7.177. The confirmed fractions exceeding $1.5\times$ and $3\times$ are 40\% and 20\%, respectively. This supports a persistent upper tail in the selected cohort but also shows attenuation after independent measurement; the complete wall-time cohort remains exploratory.

\paragraph{Small-denominator sensitivity} Requiring a baseline count of at least three leaves 658 seeds for each model. The P90 and maximum remain $31.80\times$ and $5{,}715.53\times$ for CloakFEM (DeepSeek), and $138.33\times$ and $5{,}752.72\times$ for CloakFEM (GPT). The upper tail is therefore not an artifact of one- or two-iteration baselines.

\paragraph{Variation across seeds} Table~\ref{tab:rq1-existence} preserves the source-wise view because the datasets exercise different solvers and physics distributions. Both models find high-cost cases in every included source. CloakFEM (GPT) improves the median and upper tail most clearly on Elmer and PDE-Agent, while the most extreme cases occur in PDE-Agent. NAFEMS has only six eligible seeds and is therefore evidence of coverage rather than a stable source-level estimate.

\paragraph{Seed-driven ceiling} The Elmer value $714.29\times$ is not unique
to either CloakFEM configuration. The same
\texttt{elmer\_MortarPoisson2Dsum\_ref} ceiling is reached by six of the seven
evaluated methods---random, evolutionary, OnePlusOne, ngCMA, CloakFEM
(DeepSeek), and CloakFEM (GPT)---while CMA-ES remains at $1.0\times$ on that
seed. This shared maximum reflects an unusually sensitive seed that several
proposal policies can discover; it should not be attributed solely to
agentic search.

\paragraph{Representative case studies} To make the discovered degradation concrete, we provide a dedicated three-case visualization in Appendix Figure~\ref{fig:representative-cases}. It materializes the benign and adversarial FEM inputs and reports measured solver work for complementary geometry-, coefficient-, and node-induced attacks. A 2\% annulus-radius change raises conjugate-gradient (CG) iterations from 95 to 417 ($4.39\times$); changing only the Helmholtz wave number from 12 to 23.94 raises iterations from 775 to 3,897 ($5.03\times$); and moving 58 nodes of a circular Poisson mesh raises iterations from 370 to 1,162 ($3.14\times$). The geometry and coefficient edits satisfy the bounded-edit control evaluated in RQ1.2. The displayed 58-node candidate comes from the unrestricted search and exceeds that control's per-node displacement bound; on the same seed, the bounded winner combines center and radius edits with one node movement and reaches only $1.04\times$. These case studies explain concrete mechanisms, while Tables~\ref{tab:rq1-existence} and~\ref{tab:rq1-search} provide population-level evidence.

\paragraph{Mesh-size versus conditioning amplification} Protected resolution controls do not imply fixed degrees of freedom when an editable geometry is remeshed: for example, the annulus case above changes from 2,099 to 2,512 nodes. We therefore distinguish remeshing-induced size amplification from fixed-topology coefficient or node-perturbation effects. The former remains within the attacker-controlled physical input surface, but its cost increase cannot be attributed solely to conditioning. Headline cases should be interpreted with their recorded mesh size and absolute work rather than as implementation defects inferred from ratios alone.

\paragraph{RQ1.2: Search effectiveness} The comparison includes both \tool configurations and all baselines from Section~\ref{sec:setup} under the same IR, execution path, acceptance-gate policy, manifest, and 40-query cap. It is not fully action-space matched: the LLM pipeline supports coordinated general structured patches, while the black-box baselines mutate encoded numeric leaves, supported nodes, and available material-structure entries. Results use the intersection of seeds with finite values and are reported separately for count and wall-time metrics in Table~\ref{tab:rq1-search}. Median, P90, and threshold rates summarize the best gate-accepted result per seed; Valid is the fraction of completed candidate solves accepted by the gate, $q_{>2}$ is the median first query reaching $\rho>2$ among successful runs, and Incomplete counts trajectories ending before $B=40$. Thus, the unrestricted table compares complete search systems rather than isolating the effect of an LLM proposal policy.

To separate search effectiveness from access to broad edits, Table~\ref{tab:rq1-restricted-actions} gives CloakFEM (GPT) and evolutionary search the same bounded action space on all 444 PDE-Agent seeds. We call this control \emph{strict-relative}: editable nonzero numeric values retain their sign and remain between one-half and twice the seed magnitude; original zeros remain zero; Poisson's ratio is capped at $0.49$; strings, structure, and topology cannot change; and cumulative displacement per node cannot exceed $0.05h$, where $h$ is the seed's characteristic mesh size. Panel (a) reports method-level outcomes, while panel (b) reports paired CloakFEM effects on the 411 seeds with finite results from both methods.

\begin{table*}[t]
\centering
\footnotesize
\setlength{\tabcolsep}{3.5pt}
\caption{Search outcomes under the same nominal 40-query cap.}
\label{tab:rq1-search}
\begin{tabular*}{\textwidth}{@{\extracolsep{\fill}}lrrrrrrr@{}}
\hline
Method & Median & P90 & $>1.5$ & $>3$ & Valid (\%) & $q_{>2}$ & Incomplete \\
\hline
\multicolumn{8}{l}{\emph{Count metrics ($n=680$)}} \\
Random          & 1.00 & 1.61   & 11.6\% & 4.1\%  & 25.8\% & 12 & 82 \\
Evolutionary    & 1.14 & 6.47   & 29.9\% & 17.2\% & 48.2\% & 12 & 68 \\
CMA-ES          & 1.00 & 1.86   & 12.5\% & 5.1\%  & 43.9\% & 14 & 109 \\
OnePlusOne      & 1.00 & 1.81   & 12.6\% & 5.4\%  & 63.9\% & 20 & 119 \\
ngCMA           & 1.00 & 2.17   & 16.3\% & 6.8\%  & 46.6\% & 13 & 105 \\
CloakFEM (DeepSeek) & 1.93 & 32.2   & 58.7\% & 38.8\% & 81.1\% & 6  & 118 \\
CloakFEM (GPT)      & 2.62 & 156.0  & 67.9\% & 48.2\% & 75.2\% & 4  & 176 \\
\hline
\multicolumn{8}{l}{\emph{Wall-time metric ($n=383$)}} \\
Random          & 1.18 & 3.17  & 25.6\% & 10.7\% & 40.2\% & 11   & 6 \\
Evolutionary    & 1.19 & 3.01  & 24.3\% & 10.2\% & 59.1\% & 14   & 8 \\
CMA-ES          & 1.19 & 2.81  & 23.2\% & 9.7\%  & 58.8\% & 11.5 & 8 \\
OnePlusOne      & 1.32 & 2.17  & 31.6\% & 7.3\%  & 78.6\% & 6    & 9 \\
ngCMA           & 1.45 & 2.44  & 42.0\% & 8.6\%  & 68.2\% & 10   & 7 \\
CloakFEM (DeepSeek) & 1.25 & 12.0  & 32.9\% & 20.4\% & 92.2\% & 6    & 35 \\
CloakFEM (GPT)      & 1.42 & 14.8  & 46.5\% & 26.6\% & 87.9\% & 4    & 46 \\
\hline
\end{tabular*}
\end{table*}

Table~\ref{tab:rq1-restricted-actions} shows that both strict-relative medians remain close to one, with CloakFEM (GPT) slightly higher (1.038 versus 1.017) and retaining a substantially heavier upper tail and higher success rate. On the 411 seeds finite for both methods, the geometric mean of the paired amplification ratios $\rho_{\text{CloakFEM}}/\rho_{\text{evolutionary}}$ is 1.147 (95\% CI [1.114, 1.183], $p=4.17\times10^{-22}$), and its probability of exceeding $1.5\times$ is 9.5 percentage points higher (95\% CI [$+6.6$, $+12.7$], $p=2.15\times10^{-10}$). The seed-bootstrap intervals, Wilcoxon test for amplification, and exact McNemar test for success all exclude the null for this paired control. This narrows the action-space concern for CloakFEM (GPT) versus evolutionary search on PDE-Agent, but it does not establish action-space-matched superiority over the other four baselines, for DeepSeek, or on Elmer. It also does not estimate within-seed trajectory variance because each method has one run per seed.

A conservative failure-aware sensitivity assigns $\rho=1$ whenever a method lacks a finite result and retains all 444 seeds. The geometric mean of the paired ratios remains 1.135 (95\% CI [1.105, 1.168], $p=8.89\times10^{-20}$), and the success-rate difference remains $+8.8$ percentage points (95\% CI [$+6.1$, $+11.7$] pp, exact McNemar $p=2.15\times10^{-10}$). Thus, the strict-relative advantage is not an artifact of excluding CloakFEM's additional non-finite runs.

\paragraph{Framework-level repeatability on a stratified subset}
We additionally rerun strict-relative evolutionary search with framework seeds
0, 1, and 2 on the same physics-stratified 80-seed manifest; 79 seeds are
eligible for this non-LLM baseline. Table~\ref{tab:strict-evolutionary-repeats}
shows similar medians, upper tails, and threshold rates across the three
independent trajectories. This establishes that the strict-relative
evolutionary reference is not dominated by one favorable framework seed.

\begin{table*}[t]
\centering
\footnotesize
\setlength{\tabcolsep}{4.5pt}
\caption{Strict-relative evolutionary search is stable across framework seeds
on the stratified subset.}
\label{tab:strict-evolutionary-repeats}
\begin{tabular}{lrrrrr}
\hline
Trajectory & Finite & Median & P90 & $>1.5$ & Completed $B=40$ \\
\hline
Seed 0 & 79 & 1.027 & 1.721 & 19.0\% & 53/79 \\
Seed 1 & 79 & 1.017 & 1.901 & 20.3\% & 52/79 \\
Seed 2 & 79 & 1.027 & 1.792 & 15.2\% & 52/79 \\
\hline
\end{tabular}
\end{table*}

\paragraph{Deterministic boundary-prior baseline}
To test whether the unrestricted advantage can be recovered by textbook
boundary seeking alone, we evaluate a non-LLM heuristic on the same stratified
manifest. It sweeps one numeric IR coordinate at a time over deterministic
log-scale factors from $10^{-6}$ to $10^6$ and explicitly approaches the
isotropic Poisson-ratio boundaries; it has no trajectory memory or coordinated
multi-field edits. Table~\ref{tab:boundary-prior-subset} compares it with the
saved unrestricted runs on the same manifest. The heuristic is substantially
stronger than generic evolutionary search in the upper tail, confirming that
domain knowledge explains part of the unrestricted gain. CloakFEM nevertheless
retains the larger median, P90, and success rate. This comparison isolates a
simple domain prior, not the effect of strict-relative edits.

\begin{table*}[t]
\centering
\footnotesize
\setlength{\tabcolsep}{4.5pt}
\caption{Unrestricted search on the stratified 80-seed manifest.}
\label{tab:boundary-prior-subset}
\begin{tabular}{lrrrrr}
\hline
Method & Finite & Median & P90 & $>1.5$ & Completed $B=40$ \\
\hline
Evolutionary & 79 & 1.095 & 6.745 & 26.6\% & 72/79 \\
Boundary prior (no LLM) & 79 & 1.139 & 19.379 & 39.2\% & 46/79 \\
CloakFEM (GPT) & 80 & 2.058 & 125.051 & 61.3\% & 54/80 \\
\hline
\end{tabular}
\end{table*}

\paragraph{Input similarity of discovered attacks} 
Table~\ref{tab:input-similarity} reports median overall, surface, and physical similarity together with the fraction of the full matched cohort jointly satisfying $S_{\mathrm{overall}}\geq0.95$ and each stated amplification threshold. Figure~\ref{fig:rq1-combined}(c) instead sweeps the similarity threshold from $0.85$ to $1.0$. Both use the complete 1,063-seed matched cohort as the denominator. The reference value $S_{\mathrm{overall}}=0.95$ is descriptive and was not enforced during search. Appendix~\ref{app:input-similarity} defines the structured-IR projection, component scores, and aggregation procedure.

\begin{table*}[t]
\centering

\footnotesize
\setlength{\tabcolsep}{3.2pt}
\caption{Input-space similarity and amplification of selected winners.}
\label{tab:input-similarity}
\begin{tabular}{lrrrrr}
\hline
Method & Overall & Surface & Physical & $\rho^\star>1.5$ & $\rho^\star>3$ \\
\hline
Random         & 0.997 & 1.000 & 1.000 & 11.6\% & 3.6\% \\
Evolutionary   & 0.992 & 1.000 & 0.988 & 25.7\% & 13.5\% \\
CMA-ES         & 0.997 & 1.000 & 1.000 & 14.0\% & 5.8\% \\
OnePlusOne     & 1.000 & 1.000 & 1.000 & 17.5\% & 5.4\% \\
ngCMA          & 0.994 & 1.000 & 1.000 & 21.5\% & 5.6\% \\
CloakFEM (DeepSeek) & 0.970 & 1.000 & 0.942 & 27.8\% & 17.9\% \\
CloakFEM (GPT)      & 0.969 & 1.000 & 0.949 & \textbf{38.3\%} & \textbf{27.1\%} \\
\hline
\end{tabular}
\end{table*}

The black-box baselines have the highest median similarity largely because many winners are unchanged or weakly changed and have $\rho^\star$ near one. Similarity alone is therefore not evidence of attack quality. At $S_{\mathrm{overall}}=0.95$, 38.3\% of CloakFEM (GPT) winners jointly exceed $\rho^\star>1.5$, compared with 27.8\% for CloakFEM (DeepSeek) and 25.7\% for evolutionary search. For $\rho^\star>3$, the corresponding fractions are 27.1\%, 17.9\%, and 13.5\%. Figure~\ref{fig:rq1-combined}(c) shows that this advantage persists across most of the displayed similarity range without similarity influencing the search.

\paragraph{Budget efficiency} The query logs yield $\rho^\star(b)=\max_{t\leq b}\rho(x_t)$ for $b\in\{1,\ldots,40\}$. Figure~\ref{fig:rq1-combined}(b) uses the same fixed 680-seed count cohort as Table~\ref{tab:rq1-search}; after a trajectory stops, its checkpointed incumbent is carried forward as a lower bound. It therefore complements the single $q_{>2}$ statistic without changing the cohort as the budget increases.

\FloatBarrier
\subsection{Complete RQ2 Results}
\label{app:rq2-complete}

\paragraph{Transferability} RQ2 tests whether selected expensive inputs expose numerical difficulty beyond the implementation on which they were found. The replay applies the original seed $x_0$ and exact saved CloakFEM (DeepSeek) winner $x^\star$ to independent FEM solver implementations without additional search, parameter adjustment, or LLM calls. Geometry, mesh, materials, boundary conditions, and loads or sources remain fixed within a canonical boundary-value problem. The 100-case cohort contains 45 linear-elasticity, 10 Poisson, 8 convection--diffusion, 34 Stokes, and 3 Navier--Stokes inputs, covering symmetric positive-definite (SPD), nonsymmetric, symmetric-indefinite saddle-point, and nonlinear saddle-point systems. It is a retrospective, mechanism-oriented cohort: cases must have a saved winner representable by the relevant canonical adapter, and some replay scripts order eligible cases by native amplification before applying a limit. It is therefore not a probability sample of the 1,068 seeds, and the highly unequal physics counts---especially 3 Navier--Stokes and 8 convection--diffusion cases---preclude population-level transfer claims. For a source solver $S$ and target solver $S'$, we compute
\begin{equation}
\rho_S=\frac{C(S,x^\star)}{C(S,x_0)},\qquad
\rho_{S'}=\frac{C(S',x^\star)}{C(S',x_0)}.
\label{eq:cross-solver-rho}
\end{equation}
The two ratios are normalized within solver and never mix costs from different implementations. Work is conjugate-gradient (CG), generalized minimal residual (GMRES), or minimum residual (MINRES) iterations for the corresponding linear operator, and total Newton iterations across viscosity continuation for Navier--Stokes. Threshold transfer requires both independent target solvers to have $\rho>1.5$; magnitude reproduction additionally requires each amplification to remain within a factor of two of the native ratio. Runs reaching the iteration cap are retained as lower bounds and marked separately.

\begin{table*}[t]
\centering
\footnotesize
\setlength{\tabcolsep}{2.5pt}
\caption{Transfer strength depends on the physical system.}
\label{tab:rq2-transfer}
\begin{tabular}{llrrrrrr}
\hline
Physics & Work & $n$ & Native & scikit-fem & DOLFINx & Transfer & Magnitude \\
\hline
Elasticity & CG & 45 & 10.99 & 10.25 & 10.25 & 42/45 & 42/45 \\
Poisson & CG & 10 & 2.26 & 2.21 & 2.26 & 7/10 & 7/10 \\
Convection--diffusion & GMRES & 8 & 1.80 & 1.51 & 1.51 & 4/8 & 4/8 \\
Stokes & MINRES & 34 & 5.24 & 1.53 & 1.70 & 17/34 & 1/34 \\
Navier--Stokes & Newton & 3 & 12.33 & 9.00 & 9.50 & 3/3 & 3/3 \\
\hline
\end{tabular}
\end{table*}

\paragraph{Problem-intrinsic versus implementation-specific degradation} Table~\ref{tab:rq2-transfer} reports replay outcomes and median amplification by physics and target solver; elasticity transfers consistently: 42 of 45 attacks reproduce in both the independent canonical implementations and production ElmerSolver. All 41 stronger attacks with native $\rho>2$ reproduce in Elmer. The three non-reproducing cases make only node perturbations and have weak native amplification; by contrast, every attack that drives Poisson's ratio toward the incompressible limit reproduces. Median amplification remains close across the source implementation, scikit-fem, DOLFINx, and Elmer ($10.99\times$, $10.25\times$, $10.25\times$, and $10.69\times$). This agreement identifies near-incompressible volumetric locking as a problem-intrinsic source of computational degradation rather than an artifact of one code path. Poisson presents a different pattern. Seven of ten attacks reproduce between the source DOLFINx implementation and an independently assembled scikit-fem problem, including the strongest variable-coefficient cases. In Elmer's Heat solver, three of ten remain above the $1.5\times$ transfer threshold, but only two also satisfy the factor-of-two magnitude criterion; the median falls from $2.26\times$ natively to $1.26\times$. For example, the variable-coefficient stripe case decreases from $9.49\times$ natively and in scikit-fem to $1.29\times$ in Elmer. These attacks therefore expose discretization- or implementation-specific sensitivity rather than a universally difficult PDE instance.

The expanded operator classes show that this distinction is not confined to SPD systems. Four of eight convection--diffusion attacks transfer under GMRES and preserve native-scale amplification. Seventeen of 34 Stokes attacks remain above $1.5\times$ under MINRES on both assemblies, but only one preserves the native magnitude; the original boundary-value problem therefore contributes strongly to the Stokes amplification. All three Navier--Stokes viscosity attacks transfer under independent Taylor--Hood/Newton implementations, with median amplification $9.00\times$ and $9.50\times$ versus $12.33\times$ natively. This 3/3 result provides case-level cross-implementation evidence for low-viscosity nonlinear difficulty, but the stratum is too small for a stable transfer-rate estimate.

Figure~\ref{fig:rq2-transfer} plots native source amplification against independent target amplification; marker shape identifies the target solver, the diagonal denotes equality, and the shaded band denotes agreement within $1.5\times$. The denser Stokes replay is separated into Appendix Figure~\ref{fig:rq2-transfer-stokes} to preserve point-level readability.

These selected replays show that cross-solver transfer is not binary: \tool finds both problem-intrinsic high-cost inputs that survive major implementation changes and solver- or BVP-specific cases whose threshold or magnitude weakens on replay. Detailed root-cause attribution and its evidentiary limits are reported in Appendix~\ref{app:rq2-mechanisms}.

\FloatBarrier
\subsection{Complete RQ3 Results}
\label{app:rq3-complete}

\paragraph{RQ3.1: Component ablation} RQ3 evaluates CloakFEM (GPT) configurations over all 444 PDE-Agent seeds while fixing the IR edit surface, validity gate, unaffected prompts, and maximum budget $B=40$. The reference configuration is a fresh execution of the complete Strategy--Generation--Selection pipeline and is not reused from the RQ1 campaign. The primary RQ1 run, the RQ3 reference run, and all RQ3 controls use the same IR representation without the strict-relative constraint profile; the separate Table~\ref{tab:rq1-restricted-actions} control adds that profile on top of IR. The complete reference configuration and all controls request $k=4$ candidates per move except the one-candidate-per-move control, which sets $k=1$ while retaining the Strategy Agent and its limit of three moves per iteration. Thus, every component comparison uses the fresh $k=4$ execution of the complete pipeline as its reference, while the $k=1$ arm explicitly tests candidate width. The controls remove the FEM primer, solver feedback, multi-candidate generation, learned candidate selection, or Strategy Agent parent selection; replace the Strategy Agent or Generation Agent with seeded zero-knowledge alternatives; or merge strategy and generation into one proposal call. Every arm contains 444 seed records; the reference run has 441 finite results and 249 trajectories completing all 40 queries, while the controls have between 441 and 443 finite results and between 249 and 313 completed trajectories. Figure~\ref{fig:rq3-ablation} compares how each control changes the center, upper tail, and extreme-attack frequency of finite $\rho^\star$.

\paragraph{Why the reported GPT medians differ} All four results use IR, but they summarize distinct declared runs or cohorts. Table~\ref{tab:rq1-existence} reports the primary $k=2$ run on 444 PDE-Agent count seeds (median 1.93); Table~\ref{tab:rq1-search} reports that same run on the 680-seed, seven-method matched count cohort spanning all three sources (2.62). The RQ3 complete-pipeline reference is a separate $k=4$ run on PDE-Agent (1.74), used only for component ablation. RQ3.2 is another independent $k=2$, $B=100$ trajectory and reports its fixed-cohort checkpoint lower bound at $B=40$ (2.05). Because API sampling is unseeded, differences between independent runs also include trajectory variation; these values are not interchangeable estimates from one execution.

The zero-knowledge Generation control has the largest observed decline across all five summaries in Figure~\ref{fig:rq3-ablation}: arithmetic-mean amplification falls from 127.38 to 10.25, geometric-mean amplification from 4.91 to 1.24, P90 from 188.13 to 1.46, and $\Pr(\rho^\star>100)$ from 12.0\% to 1.1\%. Removing solver feedback or the learned Selection Agent also lowers the geometric mean to 3.52 and 2.70 and the P90 to 42.09 and 17.66, respectively; the configuration without the Selection Agent nevertheless produces a higher valid-candidate fraction (85.1\% versus 74.9\%), showing that validity alone is insufficient without promising candidate selection. Replacing the Strategy Agent with zero-knowledge move scheduling yields a smaller decline across the displayed upper-tail summaries, while removing the generic FEM primer has little effect. The complete pipeline exceeds the $k=1$ and fixed-parent controls on all five displayed aggregate summaries. Their slightly higher medians (1.87 and 1.94 versus 1.74) show that the observed benefit of multi-candidate generation and adaptive parent selection lies in stronger tail outcomes rather than a higher median; the geometric mean, arithmetic mean, P90, and extreme-attack rates all favor the complete pipeline. The number of Strategy moves was not independently ablated. Because the default system contains no explicit catalog of attack mechanisms, the primer ablation measures general FEM context rather than mechanism-specific prior knowledge. Every arm has one stochastic trajectory per seed, so these comparisons describe the observed cohort rather than repeat-run stability.

\paragraph{RQ3.2: Budget sensitivity} Figure~\ref{fig:rq1-combined}(b) reports comparative efficiency through the standard budget $B=40$. An extended run applies the full CloakFEM (GPT) pipeline to all 444 PDE-Agent seeds through $B=100$, with a proportionally increased 4,500-second trajectory limit. Using checkpointed best-so-far values for incomplete trajectories, the fixed-cohort median lower bound rises from 2.05 at $B=40$ to 2.42 at $B=100$, a 17.7\% increase, while the fraction actually reaching those budgets falls from 74.3\% to 56.1\%. Search therefore continues to improve beyond the standard budget, but with diminishing returns and many incomplete trajectories. Appendix Figure~\ref{fig:rq3-budget-sensitivity} gives the complete curves, coverage, and intermediate-budget statistics.

\FloatBarrier
\subsection{Complete RQ4 Results}
\label{app:rq4-complete}

\begin{wrapfigure}[12]{r}{0.40\textwidth}
\centering
\includegraphics[width=0.96\linewidth]{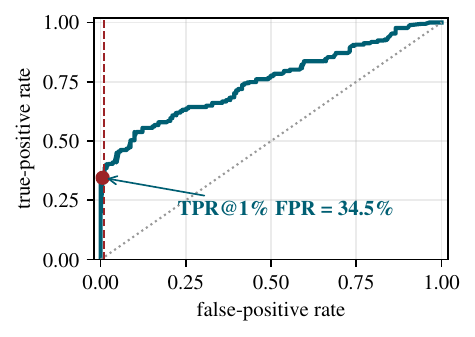}
\caption{Timeout detection rate as the false-positive allowance varies.}
\label{fig:rq4-timeout-sweep}
\end{wrapfigure}
\paragraph{RQ4.1: Runtime limits} Figure~\ref{fig:rq4-timeout} shows that a fixed limit bounds individual executions but separates benign and adversarial seeds poorly. Across 171 CG seeds, the benign- and adversarial-iteration distributions have an overlap coefficient of 0.62. Restricting false positives to 1\% detects only 34.5\% of adversarial inputs, leaving 65.5\% below the timeout threshold; detecting more adversarial inputs therefore requires rejecting a larger fraction of legitimate expensive seeds.

\paragraph{RQ4.2: Solver-side defenses} To avoid repeating the same elasticity conclusion on different cohorts, one strict-pair system-defense matrix serves as the primary RQ4.2 result. A strict pair consists of a benign seed and the higher-$\rho$ finite replayable input found by CloakFEM (GPT) or evolutionary search under strict-relative constraints. Four of 444 seeds have no finite replayable result. We retain all system classes rather than thresholding on attack success, so several rows are structure-matched controls rather than strong-attack cohorts: median native amplification is 3.06 for elasticity, 1.15 for biharmonic, 1.03 for Navier--Stokes, and 1.00 for the scalar, Stokes, and transient classes. Benign and edited inputs run consecutively in the same worker, and a defense is effective only if it combines low benign overhead, low residual amplification, and reliable completion of both runs.

Each system--defense cell applies one solver configuration to all applicable strict pairs. Linear work is the iteration count of a Krylov method---CG, GMRES, or MINRES---and nonlinear work is the cumulative number of Newton iterations. Jacobi denotes diagonal preconditioning, incomplete LU (ILU) approximates an LU factorization, algebraic multigrid (AMG) constructs a hierarchy from the discrete operator, and direct LU factorizes the matrix without Krylov iteration. Viscosity continuation solves a sequence of increasingly difficult Navier--Stokes problems ending at the target viscosity, while Backward Euler is the implicit first-order time integrator used for linear transient controls. In Table~\ref{tab:rq42-strict-followup}, $n$ is the selected strict-pair cohort, Paired counts cases in which both runs complete, benign overhead is measured against the control configuration for the same system, and Work and Time are median edited-to-benign ratios over completed pairs.

\begin{table*}[t]
\centering
\footnotesize
\setlength{\tabcolsep}{2.5pt}
\caption{Observed mitigation, benign overhead, and paired coverage.}
\label{tab:rq42-strict-followup}
\resizebox{\textwidth}{!}{%
\begin{tabular}{llrrrrr}
\hline
System class & Defense & $n$ & Paired & Benign overhead & Work ratio & Time ratio \\
\hline
Biharmonic & Stage-wise CG & 33 & 33/33 & 1.00 & 1.15 & 1.00 \\
Elasticity & CG & 45 & 45/45 & 1.00 & 2.91 & 2.88 \\
Elasticity & CG + Jacobi & 45 & 45/45 & 1.01 & 2.88 & 2.88 \\
Elasticity & GMRES + ILU & 45 & 40/45 & 10.31 & 1.33 & 1.26 \\
Elasticity & CG + AMG & 45 & 45/45 & 5.01 & 2.74 & 2.12 \\
Elasticity & Direct LU & 45 & 45/45 & 2.00 & --- & 0.98 \\
Navier--Stokes & Newton + direct & 22 & 22/22 & 1.00 & 1.33 & 1.04 \\
Navier--Stokes & Viscosity continuation & 22 & 22/22 & 2.38 & 1.03 & 1.02 \\
Poisson & CG & 79 & 79/79 & 1.00 & 1.00 & 0.99 \\
Stokes & MINRES & 45 & 45/45 & 1.00 & 1.00 & 0.99 \\
Linear transient & Backward Euler + direct & 72 & 68/72 & 1.00 & 1.00 & 0.99 \\
Nonlinear transient & Newton + direct & 37 & 37/37 & 1.00 & 1.00 & 0.99 \\
\hline
\end{tabular}
}
\end{table*}

\begin{figure*}[t]
\centering
\includegraphics[width=\textwidth]{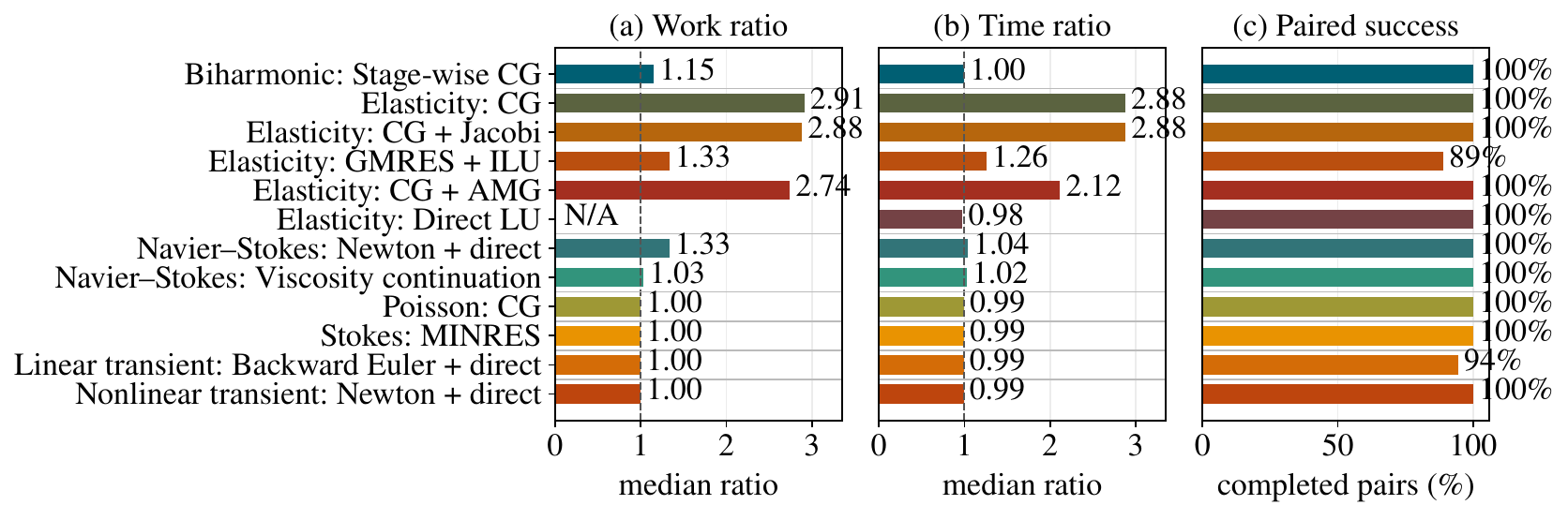}
\caption{Solver defenses trade mitigation for overhead and coverage.}
\label{fig:rq42-system}
\end{figure*}

Across 535 executed replays, 526 complete as benign--edited pairs. The nine unpaired cases comprise five elasticity GMRES--ILU failures and four linear-transient failures. Figure~\ref{fig:rq42-system} exposes the mitigation--overhead--coverage tradeoff. In the 45-seed elasticity cohort, Jacobi is effectively unchanged from CG. AMG completes all pairs but costs $5.01\times$ on benign inputs and leaves $2.74\times$ work and $2.12\times$ time amplification. GMRES--ILU lowers the observed residual amplification but incurs $10.31\times$ benign overhead and fails on five edited inputs. Direct LU completes all pairs with a $0.98\times$ edited-to-benign time ratio and $2.00\times$ benign overhead at the tested scale.

An earlier 42-case high-amplification elasticity pilot is not retained as a separate table or figure because it tests the same five configurations on a threshold-selected cohort. Its unique scale diagnostic is retained: the median number of nonzeros stored in the direct $L+U$ factors is approximately $11\times$ that of the original matrix (P90 $16.4\times$). Sparse-direct memory and factorization cost can grow superlinearly, and neither cohort covers mesh-size, fill-in, or repeated-reassembly attacks. Direct LU is therefore effective only at the evaluated scale, not a universal defense.

Viscosity continuation provides the clearest nonlinear mitigation: on 22 Navier--Stokes seeds it incurs $2.38\times$ benign overhead but reduces the cumulative-Newton ratio from $1.33\times$ under direct Newton at the target viscosity to $1.03\times$, with 22/22 paired completion. The Poisson, Stokes, biharmonic, and transient rows are structure-matched controls---reference solver configurations appropriate to those systems---rather than comparisons among alternative defenses. The implemented evaluation covers seven numerical-system classes but excludes steady convection--diffusion and Helmholtz, and it uses strict-relative selected pairs rather than the strongest unrestricted winners. RQ4 is therefore an exploratory study of the implemented defenses, not evidence for a generally effective FEM defense.

\FloatBarrier

\section{Search and IR Implementation Details}
\label{app:search-details}

Table~\ref{tab:ir-surfaces} lists the semantic edit surfaces shared across native FEM stacks. An atomic edit is $(o,p,v)$, where $o\in\{\mathrm{set},\mathrm{add},\mathrm{delete}\}$ is an operation, $p$ is a typed structured path, and $v$ is the new value when required. A candidate may combine several atomic edits. Source-specific writers apply the patch, reconstruct the native input, and restore protected blocks verbatim.

\begin{table}[h]
\centering
\footnotesize
\setlength{\tabcolsep}{4pt}
\caption{Solver-independent IR surfaces. Search can edit physical inputs but cannot modify protected solver methods, tolerances, limits, or resolution controls.}
\label{tab:ir-surfaces}
\begin{tabular}{lll}
\hline
IR component & Representative content & Search access \\
\hline
Physics & Equation or problem-class label & Read only \\
Materials & Region parameters and constitutive values & Editable \\
Conditions & Boundary conditions, loads, and sources & Editable \\
Geometry & Supported parametric dimensions & Editable \\
Mesh edits & Explicit node displacements & Editable \\
Solver state & Method, tolerances, limits, and resolution & No access \\
\hline
\end{tabular}
\end{table}

\paragraph{Archive and parent policy} The archive is initialized with $(x_0,\rho=1)$ and retains every gate-accepted execution, including non-improving states; rejected executions remain in memory but do not enter the archive. Each Strategy move may select a different archived parent. An omitted or invalid parent resolves to the highest-$\rho$ state. For baselines without learned parent selection, the framework chooses that state with probability $1-\varepsilon$ and a uniformly random archived state with probability $\varepsilon$. All reported runs use $\varepsilon=0.15$, which was not tuned by method or seed. The models do not observe $\varepsilon$ or the remaining budget.

\paragraph{Agent policies and fallbacks} The Strategy Agent proposes up to three prioritized moves. A deterministic sanitizer removes and records leaked magnitudes while preserving structural indices. For each move, the Generation Agent is instructed to produce between one and $k$ patches at different aggressiveness levels; primary searches request $k=2$. Over-complete responses are recorded rather than silently truncated, and realized candidate-set size is logged. Generation retries may use explicit rejection reasons; they consume model calls but no solver queries. If several candidates survive, the Selection Agent predicts their cost ratios and chooses one move--magnitude pair. If only one survives, selection is skipped; an invalid selection falls back to Strategy priority, Generation preference, and aggressiveness.

\paragraph{Bounded trajectory memory} For each edited target, a whole-trajectory digest records attempt count, best gate-accepted ratio, valid-but-inert edits, and gate failures, and lists at most 40 untried numeric IR paths. The loop retains eight recent concrete attempts: Strategy and Generation receive the latest six, whereas Selection receives all eight. Current-parent provenance is truncated to five edits and each archived-state summary to four. Strategy sees archived states as identifiers, ratios, short provenance, and solver-visible readings; full parent objects are supplied to Generation only for the at most three selected moves.

\FloatBarrier
\section{Three-Agent Prompts}
\label{app:agent-prompts}

\paragraph{Prompt construction} The implementation adds the current IR, solver-visible readings, search history, accepted-state frontier, and any gate-rejection feedback to each role prompt at runtime. The prompts also state that agents may edit only exposed physical inputs and supported node positions; solver methods, tolerances, limits, and resolution controls remain protected. The primary configuration supplies general FEM terminology but no catalog of attack mechanisms. The following human-readable prompts summarize the instructions given to the three agents.

\begin{prompt}[Strategy Agent]
You are the Strategy Agent in a search for runnable FEM inputs that increase the compute cost of a fixed solver. Review the current FEM problem, its cost relative to the original seed, solver-visible readings, recent attempts, gate rejections, per-target history, and the accepted-state frontier.

Propose up to three promising next moves in priority order. For each move, identify the parent state, the IR fields or physical regions to edit, the operation to apply, the qualitative direction of change, and a brief rationale. Do not choose numerical magnitudes; the Generation Agent will do that. Do not edit solver settings or unsupported fields. Avoid directions that the history shows to be invalid, exhausted, or unable to change cost.
\end{prompt}

\begin{prompt}[Generation Agent]
You are the Generation Agent. Convert each move selected by the Strategy Agent into concrete, well-typed patches to that move's parent FEM input. Use only the specified targets, operations, and qualitative direction.

For each move, generate between one and the requested number of candidates at meaningfully different magnitudes, rank them by preference and aggressiveness, and briefly state the expected effect. Use the current IR, solver-visible readings, recent attempts, per-target history, and rejection feedback when choosing values. Keep every value finite and respect the implemented geometry, constitutive, and problem-specific checks. Do not change solver methods, tolerances, limits, or resolution controls.
\end{prompt}

\begin{prompt}[Selection Agent]
You are the Selection Agent. Several concrete candidate edits have passed pre-execution validation, but none has been run. Review their parent states, diffs, intended directions, aggressiveness, the primary cost metric, solver-visible readings, and the complete search history.

Estimate the likely cost amplification of each candidate and choose exactly one for the next solver execution. Prefer candidates supported by prior improvements or promising unresolved directions, and avoid edits already shown to be inert, invalid, or exhausted. Select from the supplied candidates exactly as given; do not invent or modify an edit.
\end{prompt}

\FloatBarrier
\section{Illustrative IR Edit and Native Write-Back}
\label{app:ir-example}

Consider a two-dimensional linear-elasticity cantilever seed. Its shared IR identifies a rectangular domain, an isotropic material, a fixed condition on the left boundary, and a downward traction on the right boundary. Table~\ref{tab:ir-example} shows a candidate containing four coordinated changes rather than a single scalar edit.

\begin{table}[h]
\centering

\footnotesize
\setlength{\tabcolsep}{3.5pt}
\caption{A coordinated IR edit preserves protected structure.}
\label{tab:ir-example}
\begin{tabular}{lrr}
\hline
Editable IR leaf & Seed & Candidate \\
\hline
Geometry height & $0.20$ & $0.16$ \\
Young's modulus $E$ & $10.0$ & $18.0$ \\
Poisson's ratio $\nu$ & $0.30$ & $0.47$ \\
Right traction $t_y$ & $-1.0$ & $-1.5$ \\
\hline
\end{tabular}
\end{table}

The candidate is represented by four atomic \texttt{set} operations on \texttt{geometry.geometry\_params.height}, \texttt{materials[0].params.E}, \texttt{materials[0].params.nu}, and \texttt{boundary\_conditions[1].value[1]}. Applying the operations to a copy of the seed preserves the original object and makes the complete candidate diff explicit to the Selection Agent.

The deterministic gate first verifies that all four values are finite. It then checks that the edited rectangle has positive dimensions and can be meshed, followed by the constitutive conditions $E>0$ and $-1<\nu<1/2$. The fixed and traction boundary types are retained, so the edit changes the physical loading without changing the boundary-condition structure. After these input-decidable checks, the source-specific writer materializes the shorter geometry, assigns the edited material and traction, and executes the configured solve. Element-Jacobian and output-finiteness checks are evaluated on the materialized candidate.

For a structured benchmark, geometry fields drive Gmsh materialization while the material and boundary leaves populate the corresponding variational-form inputs. For an Elmer seed, the writer maps whichever of these leaves are exposed by that seed into its editable \texttt{.sif}/\texttt{.grd} fields. In either case, solver method, preconditioner, tolerances, iteration limits, and resolution controls are recovered from the immutable native seed rather than from model output. The example therefore illustrates both coordinated cross-component editing and the separation between the shared physical IR and source-specific protected solver syntax.

\FloatBarrier
\section{Acceptance Gate and Well-Posed PDE Subclasses}
\label{app:validity-derivation}

Compileablity and runnability are meaningful only relative to a modeled physical system, units, boundary conditions, and an application contract. The heterogeneous dataset does not expose enough common information to prove that property for every candidate. We therefore reserve \emph{gate-accepted} for the implemented predicate. For the set $\mathcal X_{\mathrm{IR}}$ of candidates representable by the supported IR and executable with the seed's protected solver configuration, define
\begin{equation}
\mathcal X_{\mathrm{gate}}
:=\mathcal X_{\mathrm{IR}}\cap
\mathcal X_{\mathrm{fin}}\cap
\mathcal X_{\mathrm{geo}}\cap
\mathcal X_{\mathrm{const}}\cap
\mathcal X_{\mathrm{solve}}.
\label{eq:gate-definition}
\end{equation}
Equation~\ref{eq:gate-definition} is the specification of the implementation, not a mathematical characterization of physical admissibility. The following paragraphs state its checks and then give well-posedness results for scalar elliptic and convection--diffusion problems, isotropic linear elasticity, and Stokes flow under explicit hypotheses.

\paragraph{Finite-data and geometry checks} Let each physical element $K$ be obtained from a reference element $\widehat K$ through a map $F_K(\widehat x;x)$ with Jacobian $J_K=\partial F_K/\partial\widehat x$. Finite element assembly requires finite input and derived quantities, while a valid fixed-topology mesh requires every element map to remain nondegenerate and preserve its reference orientation~\citep{ciarlet1978fem,hughes2000fem,strang2008analysis}. For a node-perturbed candidate, these requirements give
\begin{align}
\mathcal X_{\mathrm{fin}}
&=\{x:\text{all edited inputs and recorded solve outputs are finite}\},\\
\mathcal X_{\mathrm{geo}}
&=\{x:\det J_K(x)\det J_K(x_0)>0\quad\forall K\}.
\label{eq:validity-fin-geo}
\end{align}
The product in Equation~\ref{eq:validity-fin-geo} excludes both $\det J_K=0$ and an orientation flip without assuming that every mesh generator uses the same global sign convention. The implementation applies this sign-preservation test to explicit node perturbations with unchanged connectivity. For parametric geometry, where remeshing changes element identities, the corresponding operational condition is successful mesh generation followed by the bridge's nondegeneracy checks. These conditions make element integration well defined but do not by themselves guarantee a solvable global system.

\paragraph{Constitutive check} For three-dimensional isotropic linear elasticity, Young's modulus $E$ and Poisson's ratio $\nu$ induce shear and bulk moduli~\citep{bonet2008nonlinear,babuska1992locking}
\begin{equation}
\mu=\frac{E}{2(1+\nu)},\qquad
K=\frac{E}{3(1-2\nu)}.
\label{eq:validity-elastic-moduli}
\end{equation}
For a symmetric strain $\varepsilon$, with
$\varepsilon_{\mathrm{dev}}=\varepsilon-\tfrac13
(\operatorname{tr}\varepsilon)I$, the strain energy decomposes as
\begin{equation}
W(\varepsilon)
=\mu\lVert\varepsilon_{\mathrm{dev}}\rVert^2
+\frac{K}{2}\bigl(\operatorname{tr}\varepsilon\bigr)^2.
\label{eq:validity-elastic-energy}
\end{equation}
The gate checks $E>0$ and $-1<\nu<1/2$. The same pattern is applied by physics and semantic key across the heterogeneous IR: recognized moduli must be positive; viscosity, conductivity, density, thickness, and analogous coefficients must be nonnegative when zero represents a disabled term; yield stress and contact penalty stiffness must be positive; and unrestricted signed quantities such as sources remain subject only to finite-value checks. This semantic classification avoids incorrectly treating a Stokes viscosity named \texttt{nu} as an elastic Poisson ratio. It does not impose application-specific bounds on loads or prescribed displacements.

\paragraph{Theorem (uniformly elliptic scalar and convection--diffusion problems)} Let $\Omega\subset\mathbb R^d$ be a bounded Lipschitz domain, $V=H_0^1(\Omega)$, and
\begin{equation}
a_s(u,v)=\int_\Omega A\nabla u\cdot\nabla v+(\boldsymbol\beta\cdot\nabla u)v+cuv\,\mathrm dx.
\label{eq:scalar-weak-form}
\end{equation}
Assume $A\in L^\infty(\Omega;\mathbb R^{d\times d})$ is symmetric and uniformly elliptic, so $\xi^\mathsf T A(x)\xi\geq\alpha\lVert\xi\rVert^2$ almost everywhere for some $\alpha>0$; $\boldsymbol\beta\in W^{1,\infty}(\Omega;\mathbb R^d)$; and $c-\tfrac12\nabla\cdot\boldsymbol\beta\geq0$ almost everywhere. Then, for every $L\in V'$, there is a unique $u\in V$ satisfying $a_s(u,v)=L(v)$ for all $v\in V$, and $\lVert u\rVert_{H^1}\leq C\lVert L\rVert_{V'}$.

\paragraph{Proof} Integration by parts and the zero trace give $\int_\Omega(\boldsymbol\beta\cdot\nabla v)v\,\mathrm dx=-\tfrac12\int_\Omega(\nabla\cdot\boldsymbol\beta)v^2\,\mathrm dx$. Hence $a_s(v,v)\geq\alpha\lVert\nabla v\rVert_{L^2}^2$, which is coercive on $H_0^1(\Omega)$ by Poincar\'e's inequality. Boundedness of the coefficients makes $a_s$ continuous, so Lax--Milgram yields existence, uniqueness, and stability~\citep{ciarlet1978fem}. Setting $\boldsymbol\beta=0$ and $c=0$ covers Poisson and heterogeneous diffusion; nonzero $\boldsymbol\beta$ covers the stated coercive convection--diffusion subclass. \(\square\)

\paragraph{Theorem (isotropic elasticity: constitutive stability and well-posedness)}
Let $\Omega\subset\mathbb R^3$ be a bounded, connected Lipschitz domain whose boundary is partitioned into a relatively open Dirichlet part $\Gamma_D$ and a traction part $\Gamma_N$. Let
\[
V=\{v\in H^1(\Omega;\mathbb R^3):
\operatorname{Tr}v=0\text{ on }\Gamma_D\}.
\]
For isotropic parameters $E\in\mathbb R$ and
$\nu\in\mathbb R\setminus\{-1,1/2\}$, the constitutive energy in
Equation~\ref{eq:validity-elastic-energy} is positive for every nonzero
symmetric strain if and only if
\[
E>0,\qquad -1<\nu<\tfrac12.
\]
If these equivalent conditions hold and
$|\Gamma_D|>0$, then, for every bounded load functional
$L\in V'$, the weak displacement problem
\begin{equation}
\text{find }u\in V\text{ such that }\quad
a(u,v)=L(v)\quad\forall v\in V,
\label{eq:elastic-weak-problem}
\end{equation}
with
\[
a(u,v)=\int_\Omega
2\mu\,\varepsilon_{\mathrm{dev}}(u):
\varepsilon_{\mathrm{dev}}(v)
+K\,\operatorname{div}u\,\operatorname{div}v\,\mathrm dx,
\]
has a unique solution and
$\lVert u\rVert_{H^1}\leq C\lVert L\rVert_{V'}$.
If instead $\Gamma_D=\emptyset$, the same operator is not injective on
$H^1(\Omega;\mathbb R^3)$ because it annihilates every rigid displacement.

\paragraph{Proof}
The orthogonal decomposition
$\varepsilon=\varepsilon_{\mathrm{dev}}
+\tfrac13(\operatorname{tr}\varepsilon)I$ shows that the deviatoric and
spherical strain subspaces are independent. Equation~\ref{eq:validity-elastic-energy}
is therefore positive for every nonzero symmetric $\varepsilon$ if and only if
$\mu>0$ and $K>0$: necessity follows by choosing respectively a nonzero
traceless strain and a nonzero spherical strain, and sufficiency follows
because the two squared terms cannot vanish simultaneously. Substituting
Equation~\ref{eq:validity-elastic-moduli} gives
$\mu>0$ and $K>0$ if and only if $E>0$ and
$-1<\nu<1/2$.

Now assume these conditions and $|\Gamma_D|>0$. Orthogonality also gives
\[
a(v,v)\geq
\min\{2\mu,3K\}\lVert\varepsilon(v)\rVert_{L^2(\Omega)}^2.
\]
Korn's inequality on $V$ yields
$\lVert v\rVert_{H^1}\leq C_K
\lVert\varepsilon(v)\rVert_{L^2}$, so $a$ is coercive on $V$; it is
continuous by Cauchy--Schwarz. The Lax--Milgram theorem then gives the unique
solution of Equation~\ref{eq:elastic-weak-problem} and the stated stability
estimate~\citep{ciarlet1978fem}. Finally, if $\Gamma_D=\emptyset$, every
rigid displacement $r(x)=c+Qx$ with $Q^\mathsf T=-Q$ satisfies
$\varepsilon(r)=0$, hence $a(r,v)=0$ for all $v$. The operator consequently
has a rigid-motion kernel and uniqueness fails without a quotient-space or
additional constraint. \(\square\)

\paragraph{Theorem (steady Stokes flow)} Let $V=H_0^1(\Omega;\mathbb R^d)$, $Q=L_0^2(\Omega)$, and let the viscosity satisfy $\eta\in L^\infty(\Omega)$ and $\eta(x)\geq\eta_0>0$ almost everywhere. If the velocity--pressure spaces satisfy the divergence inf-sup condition, then for every bounded body-force functional there is a unique weak Stokes solution $(u,p)\in V\times Q$, with pressure normalized to zero mean, and the solution obeys a stability estimate in $\lVert u\rVert_{H^1}+\lVert p\rVert_{L^2}$.

\paragraph{Proof} Positive viscosity and Korn's inequality make the viscous form coercive on the divergence-free kernel. Continuity of the viscous and divergence forms together with the inf-sup condition gives existence, uniqueness, and stability by the mixed variational theorem~\citep{girault1986navierstokes}. The same conclusion holds for a conforming discrete pair when its discrete inf-sup constant is positive. \(\square\)

\paragraph{Relation of the theorems to the gate} These guarantees are not inferred from solver convergence. They apply when a candidate preserves the corresponding domain, boundary, coefficient, loading, and discretization hypotheses: uniform ellipticity and the convection--reaction sign condition for scalar problems; a positive-measure Dirichlet boundary together with $E>0$ and $-1<\nu<1/2$ for isotropic elasticity; or positive viscosity and an inf-sup-stable velocity--pressure pair for Stokes flow. They therefore cover several major operator classes in the evaluation, but not every heterogeneous seed. Helmholtz problems additionally require exclusion of resonant frequencies, and nonlinear Navier--Stokes problems require regime-dependent existence or uniqueness assumptions; for those classes we retain the local discrete criterion below rather than claim unconditional global well-posedness.

\paragraph{Local discrete solvability} After assembly, write the discrete residual as $R(u;x)=0$ and its initial linearization as
\begin{equation}
A_0(x)=\left.\frac{\partial R}{\partial u}\right|_{u=u_0}.
\label{eq:validity-linearization}
\end{equation}
If $R$ is continuously differentiable and $A_0(x)$ is nonsingular, the implicit function theorem gives a neighborhood in which a locally unique differentiable solution branch exists~\citep{krantz2002implicit}, with
\begin{equation}
\frac{\mathrm d u}{\mathrm d x}
=-A_0(x)^{-1}\frac{\partial R}{\partial x}.
\label{eq:validity-ift}
\end{equation}
Nonsingularity is thus a sufficient regularity condition for local discrete solvability, not a claim of global uniqueness or easy numerical convergence. When an assembled operator is exposed, the canonical implementation checks LU pivots relative to the operator scale. The evaluated dataset bridges do not expose a common operator or rank diagnostic, so $\mathcal X_{\mathrm{solve}}$ operationally requires the configured solve to report successful convergence and finite outputs. A rejected nonconvergent run therefore records failure of the observable completion criterion; it does not prove that the PDE is mathematically unsolvable.

\paragraph{Composed gate} For execution result $r=S(x)$, the implemented decision can be written as
\begin{equation}
V(x,r)
=V_{\mathrm{fin}}(x,r)V_{\mathrm{geo}}(x)
V_{\mathrm{const}}(x)V_{\mathrm{solve}}(r),
\qquad V(x,r)\in\{0,1\}.
\label{eq:validity-decision}
\end{equation}
Input-decidable factors are evaluated before a solver query when possible; mesh- and solve-dependent factors are evaluated after commitment. Equation~\ref{eq:validity-decision} is simply the executable gate corresponding to Equation~\ref{eq:gate-definition}; we make no proposition out of that definitional equivalence.

\FloatBarrier
\section{Dataset and Target-Solver Details}
\label{app:dataset-details}

The dataset is assembled before search and remains fixed for all methods. Table~\ref{tab:dataset-details} establishes the evaluation population before the dataset-specific processing and exclusions below. Collected counts all public specifications, while Included counts seeds admitted after confirming a usable baseline and at least one supported IR edit. Admission is independent of attack performance.

\begin{table*}[t]
\centering
\footnotesize
\setlength{\tabcolsep}{3.5pt}
\caption{The fixed manifest contains 1,068 eligible seeds.}
\label{tab:dataset-details}
\begin{tabular}{lrrp{2.6cm}p{3.0cm}}
\hline
Dataset & Collected & Included & Source representation & Target solver \\
\hline
Elmer & 1,087 & 618 & Native \texttt{.sif}/\texttt{.grd} regression cases & ElmerSolver \\
PDE-Agent & 645 & 444 & Versioned benchmark records with parametric meshes & DOLFINx 0.9 + PETSc 3.24 \\
NAFEMS & 14 & 6 & Independently reproduced benchmark IR and \texttt{.msh} meshes & scikit-fem 12.0.2 or DOLFINx scalar bridge \\
\hline
Total & 1,746 & 1,068 & \multicolumn{2}{l}{Fixed before attack search} \\
\hline
\end{tabular}
\end{table*}

\paragraph{Dataset-specific processing} For Elmer, we retain the native regression-suite \texttt{.sif}/\texttt{.grd} inputs and map editable physical fields into the shared IR; every write-back restores the original solver blocks. PDE-Agent records provide structured physical specifications, and their meshes are materialized parametrically with Gmsh before execution in DOLFINx/PETSc. NAFEMS cases use independently reproduced public benchmark specifications and versioned meshes; the supported scalar and elasticity cases execute through the corresponding DOLFINx or scikit-fem bridge. Native Elmer snapshots and all versioned IR records remain immutable during search.

\paragraph{Eligibility accounting} The manifest records execution eligibility rather than claiming that an excluded PDE is mathematically unsolvable. Of the 1,087 Elmer cases, 38 expose neither an editable parameter nor a supported editable node. Another 430 do not yield a usable baseline record in the dataset sweep: 156 are not dispatched because their mesh representation is neither \texttt{.grd}-backed nor recognized by the native-input dispatch path, while 274 are dispatched but fail. Among the latter, 250 terminate without a clean ElmerSolver completion, eight reach the 60-second baseline-sweep timeout, seven lack a usable \texttt{Mesh DB} declaration, three reference a missing native \texttt{.sif}, and six encounter mesh-generation or temporary-filesystem errors. Baseline replay additionally excludes \texttt{elmer\_MortarPoisson2D\_ref}: its initial sweep record reported 5,000 iterations, but clean replay in both model campaigns terminates with BiCGStabl numerical divergence before any attack candidate is evaluated. This outcome-independent eligibility correction leaves 618 Elmer seeds. Of the 645 PDE-Agent cases, 42 transient cases are not dispatched because their time formulation is not recognized by the transient adapter. The other 159 excluded cases are dispatched but fail during translation, mesh construction, or variational-form assembly: 118 require unsupported domain, periodic-mesh, hole, or boundary-region handling, and 41 contain missing or unhandled fields or forms, including absent \texttt{alpha}, integration domains, vertices, initial conditions, or material parameters. This leaves 444 PDE-Agent seeds. Of the 14 NAFEMS benchmarks, seven use physics unsupported by the available bridges; NAFEMS LE6 reaches the elasticity bridge but its encoded boundary conditions provide no Dirichlet constraint, leaving rigid-body modes and a singular system. The remaining six NAFEMS seeds are included.

The resulting 1,068 seeds cover biharmonic, Burgers, convection--diffusion, diffusion--reaction, large-deformation elasticity, electromagnetic wave, electrostatics, steady and transient heat, Helmholtz, small-strain isotropic elasticity, magnetostatics, Navier--Stokes, Poisson, Stokes, acoustic wave, and a residual ``other'' class used for supported Elmer regression problems that do not map cleanly to the preceding labels.

\paragraph{Fixed identity and comparability} Each manifest entry fixes the dataset source, seed identifier, physics label, editable IR surface, native writer, target solver, acceptance-gate policy, and requested cost field before any method runs. All search methods therefore operate on the same seed identities and reconstruction path. The LLM configurations and black-box baselines differ in proposal policy, not in the target solver or execution environment.

\paragraph{Metric assignment and run logging} The corrected cohort initially requests linear iterations for 469 seeds, Newton iterations for 411, and wall time for 188. A count whose baseline value is at most one cannot express meaningful amplification, so a deterministic one-time fallback selects another positive finite count or wall time before search and then locks that field. The final cohort has 685 usable count-based ratios and 383 wall-time ratios. Primary statistics use count seeds and wall-time seeds are reported separately. Count baselines below three are flagged in the result artifacts because their ratios are quantized; the small-denominator sensitivity analysis in Appendix~\ref{app:rq1-complete} reports P90 and maximum after requiring baseline count $\geq3$. The headline maximum itself has a baseline of 43 and rises to 247,367 iterations ($5{,}752.72\times$). Within each trajectory, the seed baseline is executed once and cached without consuming the query budget; each committed candidate is executed once, and the final maximum is selected from those observations without a confirmation rerun. Each run records the served model identifier, requests, failures, token and cache usage, realized candidate-set summaries, solver outcomes, and completion status. A complete method-level accounting of solver time, LLM latency, tokens, and monetary cost remains future reporting work.

\FloatBarrier
\section{Representative FEM Case Visualizations}
\label{app:representative-cases}

\begin{figure*}[!t]
\centering
\includegraphics[width=0.85\textwidth]{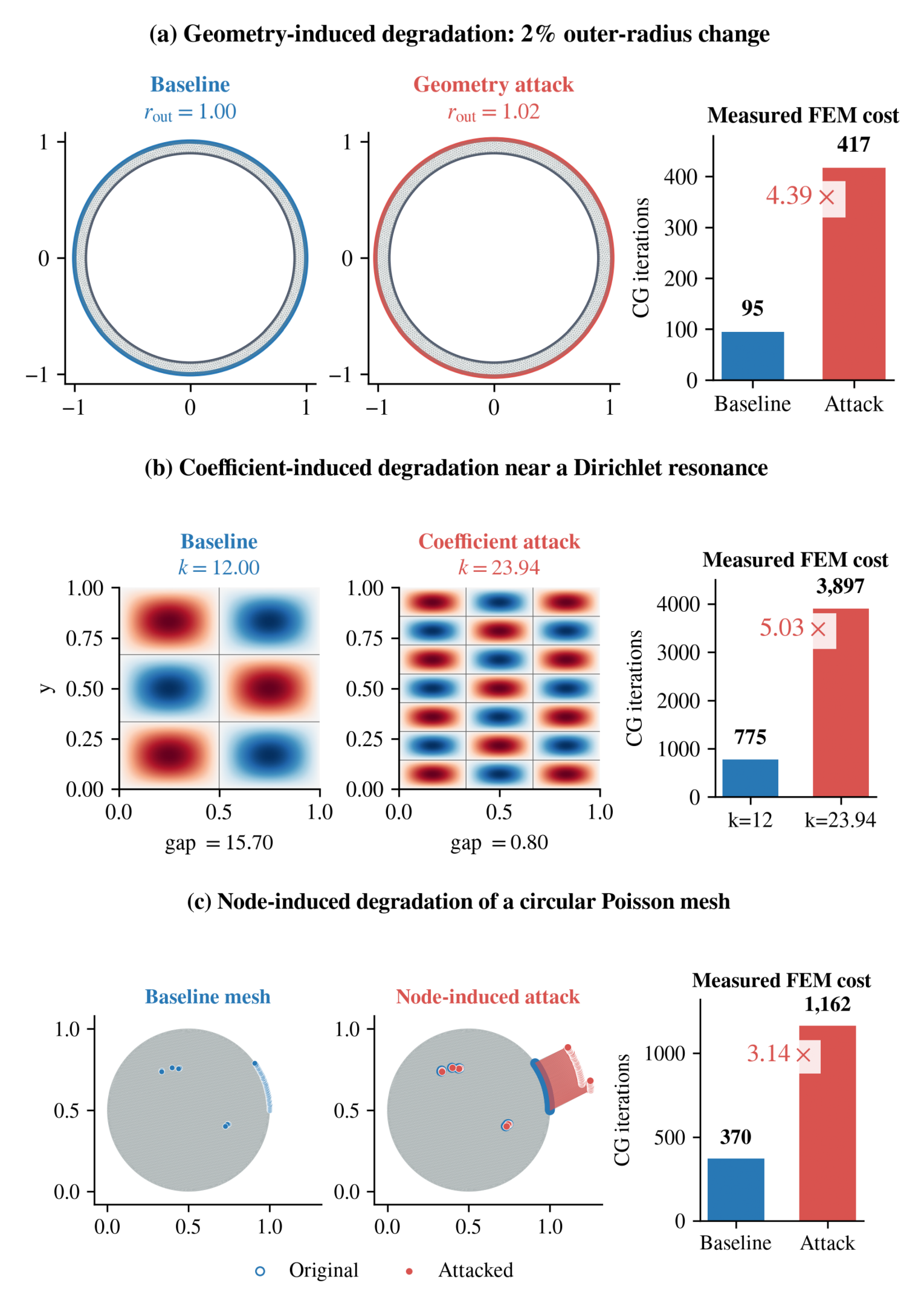}
\caption{Small FEM input edits can substantially increase solver work.}
\label{fig:representative-cases}
\end{figure*}

Figure~\ref{fig:representative-cases} presents the three complementary attack surfaces analyzed below. Each case isolates the edited input component and connects it to measured solver work.

\paragraph{Parametric geometry} In \texttt{pdeagent\_helmholtz\_annulus\_thin\_k\_high}, $r_{\mathrm{in}}=0.9$ remains fixed while $r_{\mathrm{out}}$ changes from $1.00$ to $1.02$. With $k=30$, materials, and boundary conditions fixed, the remeshed domain changes from 2,099 to 2,512 nodes and CG work rises from 95 to 417 iterations ($4.39\times$).

\paragraph{Physical coefficient} In \texttt{pdeagent\_helmholtz\_k12\_exponential\_mode}, geometry, mesh, source, and boundary conditions remain fixed while $k$ changes from 12 to 23.942385. The nearest unit-square Dirichlet spectral gap decreases from $15.70$ to $0.80$, and CG work rises from 775 to 3,897 iterations ($5.03\times$). Both this case and the geometry case satisfy the strict-relative bounds used in Table~\ref{tab:rq1-restricted-actions}.

\paragraph{Mesh nodes} In \texttt{pdeagent\_poisson\_circle\_exp\_sin}, the domain, PDE, forcing, and boundary conditions remain fixed while 58 unique nodes move. Blue outlines and red markers show benign and adversarial positions; the distortion raises CG work from 370 to 1,162 iterations ($3.14\times$). This displayed winner comes from the unrestricted search and exceeds the strict cumulative per-node limit of $0.05h=0.0003125$. The seed is included in the strict comparison, but its strict winner combines bounded center and radius edits with movement of one node and raises work from 370 to 383 iterations ($1.04\times$); it is therefore not a strict node-only analogue of the displayed case. The main-paper composite can be regenerated with \texttt{paper/figures/plot\_case\_visualizations.py}. The script materializes the benign and adversarial meshes from the stored IR records and uses the recorded iteration counts for the cost panels.

\section{Input-Similarity Measurement}
\label{app:input-similarity}

\paragraph{Matched descriptive protocol} The analysis compares the original seed $x_0$ with the saved best attack $x^\star$ for uniform random search, evolutionary search, CMA-ES, Nevergrad OnePlusOne, Nevergrad CMA, CloakFEM (DeepSeek), and CloakFEM (GPT). It uses the intersection of 1,063 seeds with records from all seven methods and is entirely post hoc: similarity neither constrains the search nor affects which winner is saved. Elmer seeds are reconstructed from the immutable native \texttt{sif0}/\texttt{grd0} stored with each search record; PDE-Agent and NAFEMS seeds are loaded from the versioned dataset. Both inputs are projected through the same structured IR used by the search implementation. Read-only context such as the materialized node count is excluded.

\paragraph{Leaf and component similarity} Let $P_0$ and $P_\star$ be the typed leaf paths in the two structured inputs. Structural similarity is their Jaccard overlap,
\begin{equation}
S_{\mathrm{top}}(x_0,x^\star)
=\frac{|P_0\cap P_\star|}{|P_0\cup P_\star|}.
\label{eq:similarity-topology}
\end{equation}
Equal categorical leaves receive score one and unequal categorical leaves score zero. For finite numeric leaves $a$ and $b$ with the same nonzero sign, the symmetric fold-change score is
\begin{equation}
s_{\mathrm{num}}(a,b)
=\frac{1}{1+\log_2\!\left(\max\{|a/b|,|b/a|\}\right)}.
\label{eq:similarity-numeric}
\end{equation}
An unchanged value scores one, a twofold change scores $0.5$, and a tenfold change scores approximately $0.23$. Zero-to-nonzero, sign, missing, and type changes score zero. Leaf scores are averaged separately for geometry ($S_{\mathrm{geom}}$), materials ($S_{\mathrm{mat}}$), boundary conditions ($S_{\mathrm{bc}}$), remaining physical configuration ($S_{\mathrm{other}}$), and categorical values ($S_{\mathrm{cat}}$).

For explicit node perturbations, displacements to the same node are accumulated and normalized by the seed's characteristic mesh size $h$. If $d_{\max}$ is the largest resulting nodal displacement, mesh similarity is
\begin{equation}
S_{\mathrm{mesh}}=\exp(-d_{\max}/h).
\label{eq:similarity-mesh}
\end{equation}
No node displacement scores one. Cases without an available $h$ retain the other components and treat mesh similarity as missing rather than silently assigning a favorable value.

The reported surface, physical, and overall scores are
\begin{align}
S_{\mathrm{surface}}
&=\operatorname{AM}(S_{\mathrm{top}},S_{\mathrm{geom}},S_{\mathrm{mesh}}),\\
S_{\mathrm{physical}}
&=\operatorname{AM}(S_{\mathrm{mat}},S_{\mathrm{bc}},S_{\mathrm{other}}),\\
S_{\mathrm{overall}}
&=\operatorname{AM}(S_{\mathrm{surface}},S_{\mathrm{physical}},S_{\mathrm{cat}}),
\label{eq:similarity-overall}
\end{align}
where $\operatorname{AM}$ denotes the arithmetic mean after omitting missing components. The decomposition prevents an unchanged mesh from concealing a large material edit, while the final scalar supports a common threshold analysis. This is only an input-space measure: it does not claim that the resulting FEM solution fields are visually or norm-wise similar, that residual or conservation error remains comparable, or that the edited input represents the same application-level task.

\FloatBarrier
\section{Numerical Mechanisms of Efficiency Degradation}
\label{app:rq2-mechanisms}

\begin{wrapfigure}[14]{r}{0.40\textwidth}
\centering
\includegraphics[width=1.10\linewidth]{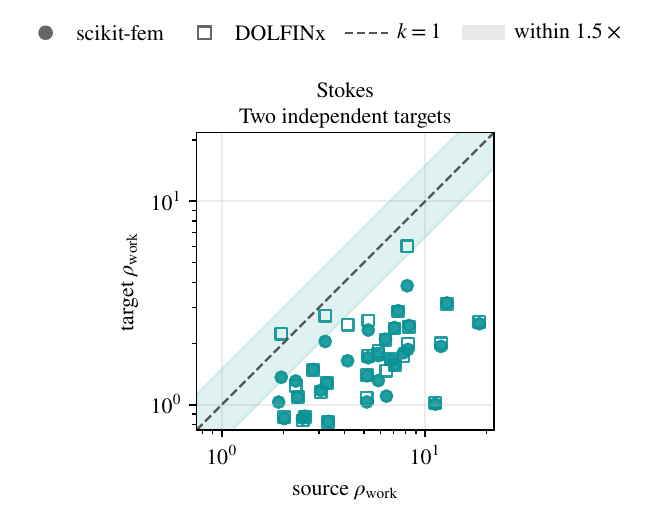}
\caption{Stokes transfer is frequent but weak.}
\label{fig:rq2-transfer-stokes}
\end{wrapfigure}
Figure~\ref{fig:rq2-transfer-stokes} reports the point-level Stokes replay used in the mechanism analysis below.

We separately analyze why selected gate-accepted inputs become expensive. Candidate mechanisms include strong material contrast, near incompressibility, mesh distortion, low-viscosity nonlinear difficulty, behavior near a limit point, contact-state changes, and sparse-graph fill-in. The discussion below is not an exhaustive taxonomy. It distinguishes direct attribution supported by exact-winner replay from a transfer pattern that does not yet identify a single causal edit.

\paragraph{Attribution protocol} Depending on target-solver support, we compare condition estimates, Krylov or Newton iterations, linear-solve time, matrix reassembly, and factorization fill between the benign seed and the exact saved winner. These diagnostics are used only after search and are not visible to the attacker. We report a diagnostic only when the replay solver exposes and records it; an unavailable spectrum, factorization statistic, or timing component remains missing rather than being inferred from solver completion. An LLM rationale is likewise not treated as mechanism evidence. A direct attribution requires an identified edit, a corresponding change in measured solver work, and reproduction in an independent implementation or a diagnostic that isolates the relevant numerical effect.

\paragraph{Near incompressibility and volumetric locking} The strongest directly supported linear mechanism is the approach to the incompressible limit in elasticity. The transferred winners move Poisson's ratio toward \(\nu=0.5\) while preserving positive elastic moduli and the other validity constraints. The cross-implementation agreement reported in Section~\ref{sec:rq2} therefore supports problem-intrinsic volumetric locking rather than a single implementation defect.

\paragraph{Coefficient contrast and discretization sensitivity} Poisson winners that increase spatial coefficient contrast exhibit the mixed transfer pattern quantified in Section~\ref{sec:rq2}. The edit therefore exposes real coefficient-induced difficulty, but its magnitude depends on discretization, assembly, and target-solver behavior.

\paragraph{Low-viscosity nonlinear difficulty} All three Navier--Stokes exact winners reduce viscosity by orders of magnitude. Two complete the independent Taylor--Hood/Newton solves with large increases in Newton work; the third reaches the Newton cap in both targets and is retained as a lower-bound transfer. Together with the ratios in Section~\ref{sec:rq2}, this supports problem-intrinsic low-viscosity nonlinear difficulty.

\paragraph{Node perturbations and mesh effects} The three elasticity attacks that fail to reproduce make only node perturbations and have weak native amplification. Their target ratios fall near one, showing that a saved geometric perturbation can exploit details of one mesh or assembly without creating a generally difficult boundary-value problem. The circular Poisson example in Figure~\ref{fig:representative-cases}(c) separately demonstrates that moving 58 nodes can raise native CG work from 370 to 1,162 iterations, but it is an unrestricted mechanism illustration rather than evidence that node-induced amplification generally transfers. Mesh distortion is therefore a plausible attack surface, while the present cross-solver evidence does not justify a universal mesh-distortion attribution.

\paragraph{Transferred operator classes without a unique attribution} The convection--diffusion and Stokes transfer reported in Figure~\ref{fig:rq2-transfer} and Figure~\ref{fig:rq2-transfer-stokes} establishes that edited boundary-value problems can remain difficult across implementations, yet the replay records do not isolate one common edit or operator diagnostic within either class. We therefore do not assign all winners in these classes to a single mechanism.

\paragraph{Mechanisms not yet established} Contact-state oscillation, behavior near nonlinear limit points, and attack-induced sparse fill remain candidate mechanisms rather than validated findings in the complete exact-winner campaigns. An earlier elasticity defense pilot records direct-factor fill with median \(11\times\) and P90 \(16.4\times\) as many nonzeros as the original matrix, but that diagnostic does not establish that the selected attacks themselves changed sparsity or mesh size. The evaluated search also locks mesh-resolution controls. Consequently, we do not infer fill-in, contact, or limit-point mechanisms from runtime alone; establishing them requires dedicated adapters and directly recorded matrix, active-set, or continuation diagnostics.

\FloatBarrier
\section{Extended-Budget Sensitivity}
\label{app:budget-sensitivity}

\begin{figure*}[!t]
\centering
\includegraphics[width=\textwidth]{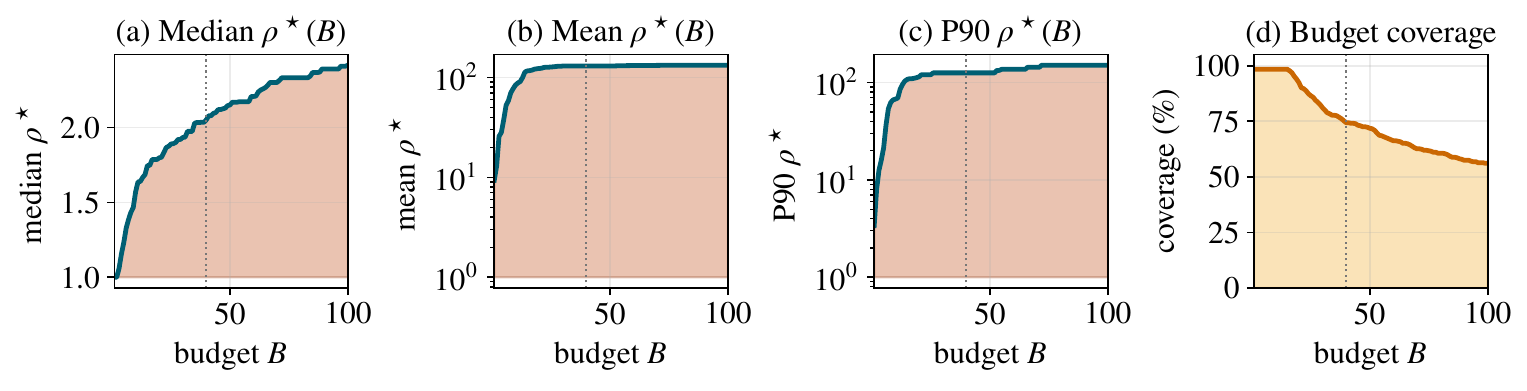}
\caption{Larger budgets improve amplification but reduce coverage.}
\label{fig:rq3-budget-sensitivity}
\end{figure*}

We extend the full CloakFEM (GPT) pipeline from the standard maximum budget $B=40$ to $B=100$ on the fixed cohort of all 444 PDE-Agent seeds. The per-trajectory limit is increased proportionally from 1,800 to 4,500 seconds. Only 249 seeds complete all 100 queries, and restricting the analysis to them would exclude many slow, high-amplification trajectories. We therefore carry each incomplete trajectory's checkpointed best-so-far $\rho^\star$ forward. Because search is monotone in its incumbent, these fixed-cohort values are conservative lower bounds rather than imputations of unobserved improvements. Figure~\ref{fig:rq3-budget-sensitivity} presents the resulting sensitivity analysis. In Table~\ref{tab:rq3-budget-sensitivity}, Reached counts trajectories that actually complete each budget, Coverage gives the corresponding cohort fraction, and the amplification summaries use the carried-forward lower bounds.

\FloatBarrier

\begin{table*}[t]
\centering
\footnotesize
\setlength{\tabcolsep}{4pt}
\caption{Larger budgets yield diminishing returns and lower coverage.}
\label{tab:rq3-budget-sensitivity}
\begin{tabular}{rrrrrr}
\hline
$B$ & Reached & Coverage & Median & Arithmetic mean & P90 \\
\hline
40  & 330/444 & 74.3\% & 2.05 & 130.06 & 125.89 \\
60  & 294/444 & 66.2\% & 2.21 & 131.45 & 136.87 \\
80  & 269/444 & 60.6\% & 2.33 & 132.39 & 150.93 \\
100 & 249/444 & 56.1\% & 2.42 & 132.54 & 150.93 \\
\hline
\end{tabular}
\end{table*}

The median lower bound increases by 17.7\% from $B=40$ to $B=100$, whereas P90 increases by 19.9\% and then plateaus after $B=80$. The arithmetic mean changes little after $B=40$ because it is already dominated by valid count-based attacks with amplification in the thousands. Median and P90 therefore provide more representative typical and upper-tail summaries. The simultaneous decline in coverage is why the extended-budget result is reported as a sensitivity analysis rather than as a direct replacement for the standard-budget comparison.

\section{Reproducibility and Reporting Details}
\label{app:reproducibility}

\noindent Code and reproduction artifacts are available at \url{https://anonymous.4open.science/r/cloakfem-5021/}.

\paragraph{Immutable seed and result identity} Every search record stores the dataset source, seed identifier, effective cost field, baseline cost, query count, best valid ratio, selected input, validity outcome, and completion status. Aggregation deduplicates reruns by seed and prefers the finite record with the largest completed query count.

\paragraph{Execution boundary and randomness} The LLM proposes structured edits, while source-specific writers, pre-execution checks, solver invocation, post-solve validity, query accounting, checkpointing, and final ranking remain deterministic framework operations. Framework-controlled sampling uses NumPy's generator with replicate seed 0, including the fixed $\varepsilon=0.15$ archive policy and seeded black-box proposals. The model APIs are called at temperature 0.7 with their standard sampling behavior. Numerical-library threads are fixed to one, and the software versions, hardware, and timeout limits are reported in Section~\ref{sec:setup}.

\paragraph{Generated artifacts} The main evaluation figures are regenerated by:\par\noindent\path{paper/figures/plot_evaluation_figures.py}\par\noindent The joint similarity analysis uses:\par\noindent\path{experiments/measure_attack_similarity.py}\par\noindent\path{paper/figures/plot_similarity_tradeoff.py}\par\noindent Representative cases use:\par\noindent\path{paper/figures/plot_case_visualizations.py}\par\noindent The extended-budget fixed-cohort analysis uses:\par\noindent\path{experiments/aggregate_budget100_sensitivity.py}\par\noindent The corresponding CSV, JSON, JSONL, PNG, and PDF outputs are retained with the experiment results. Stochastic uncertainty is reported across seeds; unless stated otherwise, each seed--method pair has one replicate and should not be interpreted as estimating between-run model variance. Table~\ref{tab:strict-evolutionary-repeats} is the exception: it reports three framework seeds for evolutionary search on a stratified subset, but no successful repeated model trajectories.

\section{Extended Background and Related Work}
\label{app:related-work}

\subsection{Finite-Element Inputs and Numerical Cost}

FEM discretizes a partial differential equation over a mesh into a sparse algebraic system, or, for nonlinear and transient problems, a sequence of such systems~\citep{hughes2000fem,strang2008analysis}. The mesh and element type determine the sparsity pattern, while geometry, material coefficients, constitutive laws, boundary conditions, and loads determine the entries and right-hand side. Consequently, two accepted inputs of comparable size can induce very different numerical work.

For sparse direct methods, factorization cost is dominated by fill-in and therefore depends on the sparsity pattern and elimination ordering rather than only on the number of nonzeros~\citep{george1981sparse,davis2006direct}. For Krylov methods, convergence depends on the operator spectrum and the effectiveness of the preconditioner; coefficient jumps, anisotropy, and mesh distortion can degrade both~\citep{saad2003iterative,trefethen1997numerical}. Nonlinear solves add Newton iterations, line-search backtracking, and step cuts, whose counts depend on constitutive nonlinearities and penalty stiffnesses. Transient solvers additionally couple work to stability limits and adaptive time stepping, while remeshing or adaptive refinement can change the degrees of freedom as a function of the submitted model.

Well-posedness and good conditioning are distinct. Near-incompressible elasticity, for example, is a meaningful physical regime but can produce volumetric locking in standard displacement formulations~\citep{babuska1992locking}; similarly, ordinary heterogeneous materials can create strong coefficient contrast. An accepted model can therefore converge to a valid solution while consuming substantially more work. This distinction motivates our acceptance gate: it rejects malformed, unsupported, or nonconvergent candidates, but does not equate acceptance with easy numerical solution or application-level realism. The distinction is operationally important for shared HPC queues, simulation services, and solver-backed endpoints, where the party specifying a model may not bear its full execution cost~\citep{baratta2023dolfinx,elmer2013,balay2019petsc}.

\subsection{Efficiency Degradation as an Attack Surface}

Algorithmic-complexity attacks established the general pattern of a legitimate input that produces the expected output while forcing disproportionate work. Hash-collision attacks exploit worst-case dictionary behavior~\citep{crosby2003dos}, and regular-expression denial of service drives backtracking matchers into pathological execution~\citep{staicu2018freezing,davis2018redos}. These attacks differ from request-flooding denial of service: amplification arises from the computational structure of an individual accepted input.

Related attacks on learned systems maximize energy or latency, defeat early-exit mechanisms, delay end-of-sequence generation, or inflate post-processing work~\citep{shumailov2021sponge,hong2021sloth,chen2022nmtsloth,feng2024llmeffichecker,dong2025engorgio,shapira2023phantom}. Their mechanisms are typically data-dependent branches or variable-length computation inside a learned pipeline. CloakFEM instead targets convergence and discretization behavior in deterministic numerical software. The attacker submits one gate-accepted model, cannot alter protected solver controls, and optimizes measured work rather than prediction error. Work on adversarial scientific machine learning attacks learned surrogates such as neural operators and evaluates accuracy degradation~\citep{adesoji2022fno}; our victim, objective, and failure mode are therefore different.

Numerical software also differs from architectures with an explicit maximum number of exits, tokens, or retained detections: solver work depends on the submitted problem and convergence trajectory rather than a single fixed branch count. Cost can rise through several mechanisms, including remeshing and larger systems, degraded conditioning at fixed topology, or slower nonlinear and transient convergence. We report mesh size, absolute work, and solver outcomes where available so that these mechanisms are not conflated with crashes, malformed inputs, or silent accuracy failures.

\subsection{Code Agents and LLM-Guided Optimization}

Language models for code progressed from function completion~\citep{chen2021codex,austin2021program} to agents that edit repositories and use execution feedback~\citep{jimenez2024swebench,yang2024sweagent,wang2025openhands}. A related line places a language model inside an externally evaluated search loop: FunSearch and AlphaEvolve evolve artifacts against deterministic evaluators~\citep{romeraparedes2024funsearch,novikov2025alphaevolve}, while OPRO and Evolution of Heuristics use language models as optimizers or heuristic designers~\citep{yang2024opro,liu2024eoh}. These systems show that semantic proposal generation can complement conventional derivative-free search.

CloakFEM follows this outer-loop pattern but changes both the search object and the objective. Its candidates are typed edits to heterogeneous FEM specifications rather than free-form programs or flat vectors, and each evaluation requires a solver execution. Strategy, generation, and selection agents propose targets, instantiate magnitudes, and choose candidates, while representation checks, native write-back, acceptance decisions, archive updates, and query accounting remain deterministic. The model does not observe assembled matrices, spectra, preconditioner state, or hidden solver configuration. This separation makes candidate edits auditable and allows the system to search semantically without granting direct access to the numerical mechanism being stressed.

The direction of optimization also differs from work that seeks more correct, faster, or better-converged artifacts: CloakFEM searches for gate-accepted, runnable inputs that increase the cost of an unchanged solver. We therefore compare it with random, evolutionary, CMA-ES, OnePlusOne, and ngCMA search under a common solver-query cap. These baselines are derivative-free proposal policies rather than agents~\citep{hansen2001cmaes}; the comparison tests whether semantic access improves evaluator-query efficiency, not whether all methods have equal proposal-side compute or identical action spaces.

\end{document}